\documentclass[sigconf,nonacm]{acmart}

\AtBeginDocument{  }

\renewcommand\footnotetextcopyrightpermission[1]{}
\setkeys{acmart.cls}{balance=false}

\makeatletter
\patchcmd{\@mkauthors@iii}
  {\par}
  {}
  {}{\ClassWarning{acmart}{Could not compact the first author in a group}}
\patchcmd{\@mkauthors@iii}
  {\par}
  {\space}
  {}{\ClassWarning{acmart}{Could not compact subsequent authors in a group}}
\makeatother

\usepackage{tikz}
\usepackage{amsmath}
\usepackage{xspace}
\usepackage{ifthen}
\usepackage{enumitem}
\setlist[itemize]{noitemsep, topsep=0pt, partopsep=0pt, left=0.5em}
\usepackage{graphicx}
\usepackage[caption=false]{subfig}
\usepackage{pifont}
\usepackage{makecell}
\usepackage{multirow}
\usepackage{microtype}
\microtypecontext{spacing=nonfrench}
\usepackage{booktabs}
\usepackage{threeparttable}
\usepackage{xurl}
\usepackage{xcolor}
\usepackage{subcaption}
\usepackage{marvosym}

\definecolor{lightgray}{gray}{0.9}

\newcommand{\solidnum}[1]{  \ifcase#1\relax\or
  \ding{182}\or
  \ding{183}\or
  \ding{184}\or
  \ding{185}\or
  \ding{186}\or
  \ding{187}\or
  \ding{188}\or
  \ding{189}\or
  \ding{190}\or
  \ding{191}\else
  \textbf{#1}\fi
}

\usepackage{tcolorbox}
\tcbuselibrary{skins, breakable}

\newcounter{finding}

\newtcolorbox{findingbox}[1][]{
  enhanced,
  breakable,
  colback=black!3,
  colframe=black!30,
  boxrule=0.2pt,
  leftrule=3pt,
  left=8pt,
  right=8pt,
  top=6pt,
  bottom=6pt,
  sharp corners,
  #1
}

\newcommand{\finding}[2][]{  \refstepcounter{finding}  \begin{findingbox}[#1]
    \textbf{Finding~\Roman{finding}.} \textit{#2}
  \end{findingbox}
}

\newtcolorbox{promptbox}[1]{
  breakable,
  colback=gray!5!white,
  colframe=gray!75!black,
  fonttitle=\bfseries,
  title=#1,
  sharp corners=southwest,
  boxrule=1pt,
  arc=4pt,
  left=6pt,
  right=6pt,
  top=6pt,
  bottom=6pt
}

\newcommand{\ttbf}[1]{\noindent\textbf{#1.}\ }

\newcommand{\fudanmark}{\textsuperscript{\textdagger}}
\newcommand{\siimark}{\textsuperscript{\textdaggerdbl}}
\newcommand{\corrmark}{\textsuperscript{\Letter}}
\begin{document}

\title{Rethinking MCP Security: A Large-Scale Study of Runtime MCP Servers and Security Scanner Reliability}

\author{\mbox{Pei Chen\fudanmark}}
\author{\mbox{Baichao An\fudanmark}}
\author{\mbox{Mengying Wu\fudanmark}}
\author{\mbox{Binwang Wan\fudanmark}}
\author{\mbox{Geng Hong\fudanmark\corrmark}}
\author{\mbox{Jinsong Chen\fudanmark}}
\author{\mbox{Xudong Pan\fudanmark\siimark}}
\author{\mbox{Jiarun Dai\fudanmark}}
\author{\mbox{Min Yang\fudanmark\corrmark}}
\affiliation{  \institution{\fudanmark Fudan University, \siimark Shanghai Innovation Institute\\
  \{peichen19, bcan20, ghong, xdpan, jrdai, m\_yang\}@fudan.edu.cn, \{wumy21, bwwan25, jschen23\}@m.fudan.edu.cn\\
  \Letter Co-corresponding authors}
  \city{Shanghai}
  \country{China}}
\renewcommand{\shortauthors}{Chen et al.}

\begin{abstract}

The Model Context Protocol (MCP) has rapidly established itself as a standard interface for enabling LLM-based agents to interact with external tools and services. As MCP servers are increasingly entrusted with security-sensitive operations, understanding their real-world risks has become critical.
In practice, due to the absence of large-scale runtime MCP servers, such understanding largely relies on security scanners applied to a small number of cases, yet the reliability of these assessments remains unclear.

In this study, we revisit how MCP security is measured.
We present MCPZoo, the largest collection of MCP servers for dynamic analysis to date.
MCPZoo is constructed through a multi-agent framework for transforming in-the-wild static repositories into dynamic services.
The framework emulates how human experts build, diagnose, and iteratively repair deployment and runtime defects by combining environment inference with feedback-driven refinement.
To ensure practical interactivity at runtime, the servers are validated via real protocol interactions.
As a result, MCPZoo contains 64,611 unique MCP servers (113,927 in total), with more than 37,288 supporting dynamic analysis.
Leveraging MCPZoo, we conduct the first ecosystem-scale measurement of MCP servers and the scanners that analyze them.
While existing scanners report that 96.89\% of servers are risky, we find that these signals are unreliable.
In particular, manual validation shows that less than 50\% of sampled alerts are true positives, and scanner outputs exhibit clear inconsistency across scanners.
Overall, MCPZoo enables large-scale, reproducible measurement of MCP server security and exposes limitations of current scanning practices.
We further release a public query interface to support practical risk assessment of MCP servers.

\end{abstract}


\maketitle

\section{Introduction}

As Large Language Models (LLMs) evolve into autonomous agents, their ability to act in the real world increasingly relies on structured access to external tools.
The Model Context Protocol (MCP) provides a standardized interface for such interactions, leading to a rapidly growing ecosystem of MCP servers.
These servers expose powerful capabilities, including file access, network communication, and system-level operations, making them a critical control point for real-world actions.
This shift introduces new security risks. Unlike traditional components, MCP servers directly execute actions on behalf of agents, so a single unsafe interaction may trigger unintended data flows or system-level effects. While prior work has identified risks such as prompt injection and unsafe tool invocation~\cite{wang2025mpma,song2025protocolunveilingattackvectors,zhao2025mindserversystematicstudy}, it remains unclear how these risks manifest in the real-world MCP ecosystem.

\ttbf{Problem}
Despite these concerns, we still lack a clear understanding of how secure the MCP ecosystem actually is in practice. Existing evidence is largely derived from isolated case studies or automated scanners, without systematic validation at scale. As a result, it remains unclear whether reported risks reflect real vulnerabilities, or artifacts of measurement methods.

At the core of this gap lies a fundamental limitation: \textit{MCP servers are not readily measurable at scale}. First, many security-relevant behaviors only emerge during runtime interactions between agents and servers, making static analysis insufficient. Second, real-world MCP servers are difficult to deploy due to complex dependencies, incomplete configurations, and protocol-specific assumptions. These challenges prevent large-scale, realistic evaluation.

\ttbf{MCPZoo}
To address this gap, we construct MCPZoo, a large-scale dataset designed to support dynamic analysis of real-world MCP servers at runtime.
MCPZoo is built through a fully automated multi-agent framework that emulates expert workflows for transforming fragmented MCP server repositories collected from multiple public sources into analyzable runtime states.
The framework continuously collects MCP server projects, infers their runtime environments, and iteratively builds, deploys, and validates servers through standardized protocol-level interaction checks.
By explicitly accounting for deployment and runtime defects that arise in real-world MCP servers, this framework systematically handles issues such as dependency mismatches, implicit configuration assumptions, and failures in protocol initialization and interaction, thereby enabling servers to meaningfully expose their behavior under realistic execution conditions.

In total, MCPZoo collects 64,611 MCP server projects (113,927 in total) from multiple public markets, among which 37,288 servers support dynamic analysis.
This provides a realistic execution environment for ecosystem-scale measurement of MCP servers.

Leveraging MCPZoo, we revisit two fundamental questions about MCP security.

\noindent\textbf{How does the MCP ecosystem behave in practice?}
We conduct a systematic characterization of the MCP ecosystem.
We find that the apparent scale of the ecosystem is largely inflated by duplication, arising through forks, copies, or mirrors, and the most frequently reused template appears 537 times. We further observe that deployment support is limited across the ecosystem. Specifically, 78.6\% of servers provide no deployment artifacts~(i.e., Dockerfiles), and only 19.6\% of projects with complete Dockerfile, README, and configuration artifacts run successfully without modification.
Leveraging MCPZoo, we achieve a global deployment success rate of 57.7\%, enabling runtime analysis at unprecedented scale. For servers that support interaction, tool exposure varies widely. While most servers expose fewer than 10 tools, the largest server exposes 8,816 tools, and 37.6\% of tools provide high capability such as command execution, file modification, or data transmission.

\noindent\textbf{Can existing scanners reliably measure MCP security?}
We perform a large-scale empirical evaluation of eight popular MCP security scanners on MCPZoo, covering static, dynamic, and hybrid analysis approaches.
At first glance, their reports appear alarming: 96.89\% of interactable servers are flagged as risky by at least one scanner.
However, deeper analysis reveals that this apparent prevalence is misleading.

A closer examination reveals three key limitations.
First, through manual sampling of scanner reports, we find that their average precision is only 45.53\%, with the worst-performing scanner achieving as low as 10.40\%. Second, scanner outputs exhibit substantial inconsistency: the average pairwise Jaccard similarity between scanner-reported server sets is only 15.66\%. Third, using a CVE-based ground-truth dataset constructed from 10 real-world vulnerabilities, we observe that scanners detect only 24.17\% of known vulnerable cases, indicating limited recall on confirmed issues.
These limitations stem from systematic issues, including reliance on metadata patterns, incomplete runtime interaction, and heuristic or LLM-based inference not grounded in exploitable behavior.
Together, these findings shift the conclusion from \textit{MCP servers are unsafe} to a more critical insight: \textit{current MCP security scanners are not yet reliable enough to support ecosystem-level security claims}.

As a supporting service, we provide a public query interface that maps MCP servers to multi-scanner reports, cross-scanner agreement, and validation status when available, helping users to assess the potential risks of MCP servers while supporting ecosystem transparency and self-assessment across the MCP community.
We hope these efforts contribute to a more reliable and trustworthy foundation for MCP security in practice.

\ttbf{Contribution}
We make the following contributions:

\begin{itemize}
	\item We present MCPZoo, a large-scale dataset that supports dynamic analysis of real-world MCP ecosystem.
	\item We design a fully automated, multi-agent framework that builds, deploys, and verifies  MCP servers at scale.
    \item Using MCPZoo, we conduct a characterization of the MCP server ecosystem, revealing extensive duplication and weak deployment practices.
	\item We perform a comprehensive reliability evaluation of MCP security scanners, showing that they are not yet reliable enough for ecosystem-level security assessment.
\end{itemize}

\section{Background}

\subsection{Model Context Protocol}

Model Context Protocol (MCP) is an open standard designed to provide a unified interface for LLM–driven agents to access external capabilities. Its core goal is to establish a general and extensible interface layer between models and external systems.

MCP system involves three roles: (i) an \textit{LLM-based agent} that performs reasoning and decision making, (ii) an \textit{MCP client} that manages protocol-level communication, and (iii) one or more \textit{MCP servers} that expose available capabilities and execute concrete operations.
Among these, \textit{MCP servers} play an important role, as they directly implement and enforce the actions available to agents.
Vulnerabilities at the server level can therefore damage the security of the entire agent system, motivating us to conduct a focused and systematic study of real-world \textit{MCP servers}.

MCP servers can be accessed through different interaction mechanisms, including \textit{stdio}, \textit{Streamable HTTP}, and \textit{Server-Sent Events (SSE)}, which differ in communication patterns and deployment settings.
These mechanisms are widely used in practice and introduce challenges for unified interaction and large-scale analysis.

\subsection{Security Risks in MCP Servers}

The MCP ecosystem introduces a set of security risks, such as prompt injection~\cite{wang2025mpma}, tool poisoning~\cite{zhao2025mindserversystematicstudy}, etc.
Since MCP servers directly mediate agent actions, vulnerabilities at the server level can readily escalate into system-level consequences. In this work, we consider adversaries who can influence agent inputs or interact with MCP servers via exposed interfaces (e.g., tool invocation or external data sources), potentially triggering unsafe tool execution or unintended data flows. We focus on risks arising from server implementations and their runtime interactions with agents, rather than attacks on the underlying LLM or client infrastructure.

However, existing analyses remain limited in scale. As summarized in Table~\ref{tab:prior_study_scale}, prior static studies cover server repositories, and the two largest reported counts are not deduplicated server populations: Hou et al.~\cite{hou2025modelcontextprotocolmcp} aggregate platform-reported catalog sizes across MCP collections, while Li et al.~\cite{li2025understandingsecurityissuesmodel} collect registry entries and note that some links refer to the same repository. while dynamic experiments rely on small or manually curated MCP server sets~\cite{song2025protocolunveilingattackvectors,noever2025servantstalkerpredatorhonest,wang2025mcptoxbenchmarktoolpoisoning}.
Even the largest existing dynamic study analyzes only 1,360 servers and validates them through limited interactions, restricting the coverage of runtime behavior~\cite{zhao2025mindserversystematicstudy}.
As a result, most analyses remain unable to validate runtime security properties at large scale, making it difficult to conduct representative and reproducible ecosystem security measurements.

\begin{table}[t]
\centering
\caption{Comparison of Existing Studies on MCP Servers.}
\label{tab:prior_study_scale}
\resizebox{\linewidth}{!}{
\begin{tabular}{lcc}
\toprule
\textbf{Work} & \textbf{\# Static Servers} & \textbf{\# Dynamic Servers} \\
\midrule
Hou et al.~\cite{hou2025modelcontextprotocolmcp}       & 105,378    &  -- \\
Li et al.~\cite{li2025understandingsecurityissuesmodel} & 67,057 & --\\
Stein et al.~\cite{stein2026aiagentsusedevidence} & 19,338 & --\\
Guo et al.~\cite{guo2025mcpecosystem}  & 17,630  & --\\
Zhao et al.~\cite{zhao2025mindserversystematicstudy} & 12,700 & 1,360 \\
Hasan et al.~\cite{hasan2026modelcontextprotocolmcp} & 1,899 & 83 \\
Noever~\cite{noever2025servantstalkerpredatorhonest}    & -- & 95\\
Tiwari et al.~\cite{tiwari2025modelcontextprotocolvision} & -- & 91 \\
MCPTox~\cite{wang2025mcptoxbenchmarktoolpoisoning}    & -- & 45\\
Song et al.~\cite{song2025protocolunveilingattackvectors} & -- & 13\\

MCPSecBench~\cite{yang2025mcpsecbenchsystematicsecuritybenchmark} & -- & 5\\

\midrule
\textbf{This work (Total)}       & 113,927   & 67,207 \\
\textbf{This work (Unique)}  & 64,611 & 37,288 \\
\bottomrule
\end{tabular}
}
\end{table}

\section{MCPZoo}

In this section, we introduce the construction of \textbf{MCPZoo}, a large-scale dataset, supporting framework for dynamic analysis of real-world MCP servers.

\subsection{Challenges \& Key Ideas}

To enable effective large-scale security analysis of the MCP ecosystem, our design is guided by three key goals:
\textit{large-scale coverage} to ensure representativeness, \textit{dynamic analysis support} to capture runtime-dependent security issues, and \textit{automated construction and deployment} to enable scalable, consistent, and reproducible evaluation.
However, achieving these goals in practice is non-trivial and gives rise to several fundamental challenges.

\textbf{\textit{Challenge 1: }}
Due to varying development quality and frequent updates, MCP server repositories may provide missing, incomplete, or outdated deployment documentation that no longer aligns with the current code.
As a result, deployment instructions in README files are often insufficient or incorrect, leading to failed deployment.
In practice, we observe that deployment entry points can be inferred by reusing prior deployment patterns.
Therefore, we first summarize common MCP deployment patterns from successfully deployed MCP servers and use them to guide inference by consulting both source codes and available documentation, with execution feedback further used to refine decisions when deployments fail.

\textbf{\textit{Challenge 2: }}
When faced with long contexts, LLM tends to overlook important details and produce biased or incomplete feedback~\cite{liu2024lost,gao2024insights}.
As a result, directly feeding LLMs with whole repositories and long deployment logs becomes increasingly ineffective as code and error records accumulate during iterative deployment.
Therefore, we form a minimal but sufficient evidence set for LLM input by emphasizing structural features and high-level error cues.
Specifically, the evidence includes structural features such as file trees, key configuration files, and distilled conclusion-level statements extracted from error logs, with few-shot examples guiding the common failures.

\textbf{\textit{Challenge 3: }}
Semantic inspection of deployment artifacts (i.e., Dockerfiles) alone is insufficient to assess whether an MCP server can support dynamic analysis in practice. Many failures only occur during runtime execution and interaction.
We observe that deployment success must be validated through protocol behavior.
Accordingly, we construct dedicated agents to perform MCP protocol–based interactions and validate deployments through direct interaction with a real MCP client.
Only servers that respond correctly to expected interactions are considered to support dynamic analysis in practice.

\begin{figure*}[t]
    \centering
    \includegraphics[width=0.9\linewidth]{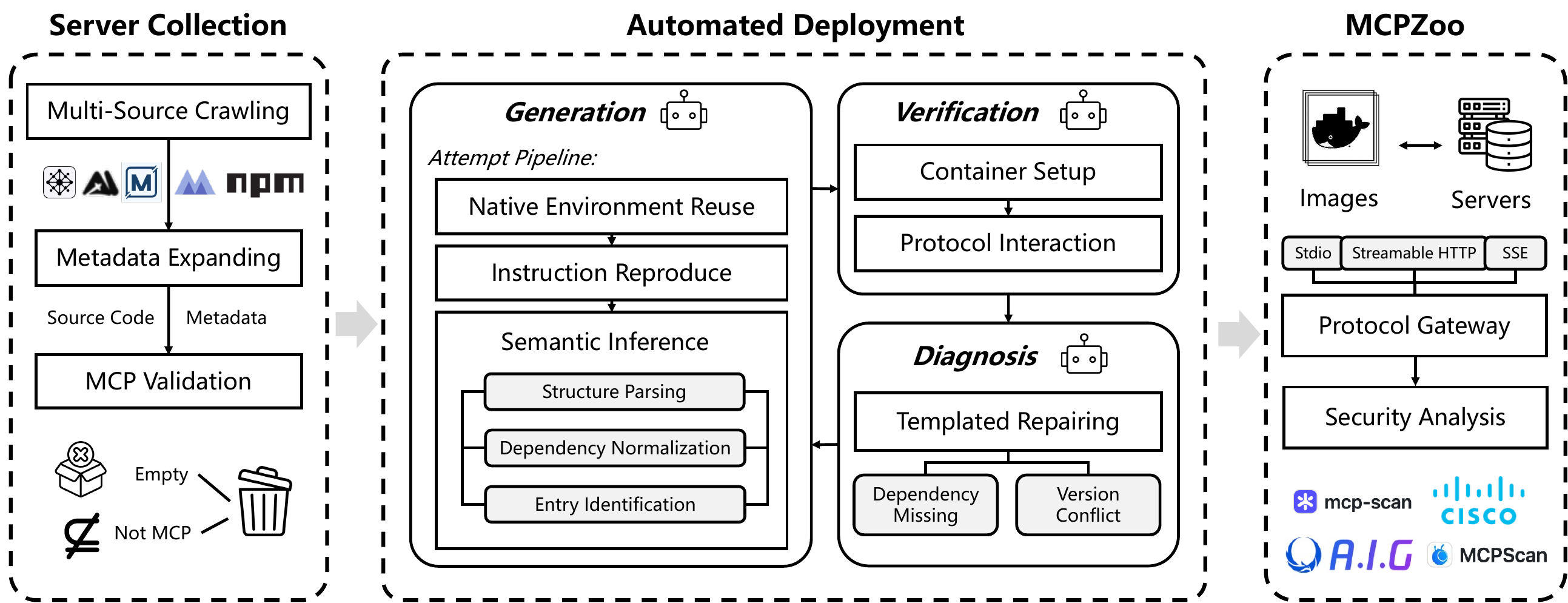}
    \caption{Overview of the MCPZoo Construction and Analysis Workflow.}
    \Description{Workflow diagram showing the main stages of MCPZoo construction and analysis.}
    \label{fig:overview}
\end{figure*}

\subsection{Overview}
To systematically construct a large-scale testing environment that supports dynamic analysis of MCP servers, we design and implement an end-to-end framework for building MCPZoo, as illustrated in Figure~\ref{fig:overview}.
The framework automatically collects MCP server projects from multiple public markets and transforms them, through an automated multi-agent system, into server instances suitable for runtime analysis under realistic execution conditions.

The three specialized agents emulate expert workflows to build a server.
The \textit{Generation Agent} is responsible for producing deployment configurations, i.e., Dockerfiles, from raw MCP server project sources.
Given a generated Dockerfile, the \textit{Verification Agent} instantiates the execution environment, launches the server, and performs multi-level runtime checks, including protocol-level interaction validation.
When verification fails, a \textit{Diagnosis Agent} analyzes error signals collected at different stages, such as build failures, startup errors, or interaction-level issues.
Based on these signals, the agent summarizes adjustment strategies and feeds them back to guide the next round of generation.
This process iterates until the server supports real dynamic analysis under runtime execution conditions or a predefined retry limit is reached.
This agent-based design enables scalable and parallel processing of MCP server projects while maintaining reliable and consistent validation across the dataset.

\subsection{MCP Server Collection}\label{sec:server_collection}

MCPZoo starts with collecting MCP server projects from multiple publicly accessible MCP markets to capture a broad and representative view of the real-world ecosystem.

The MCP protocol does not require servers to be published through a centralized registry. Although an official registry has been established by the protocol proposer Anthropic, it currently indexes only part of the MCP servers compared to the wild~\cite{anthropic-mcp-donation,mcp-registry}. As a result, MCP servers are mainly distributed across multiple third-party community markets, leading to a decentralized and fragmented publication landscape.
Specifically, our sources include dedicated MCP markets such as MCP Store~\cite{mcpstore}, MCP World~\cite{mcpworld}, MCP Market~\cite{mcpmarket}, MCP Servers Repository~\cite{mcprepository}, AIBase MCP~\cite{aibasemcp}, Pulse MCP~\cite{pulsemcp}, MCP.so~\cite{mcpso}, Smithery~\cite{smithery}, as well as general-purpose package ecosystems including NPM~\cite{npm_registry} and PyPI~\cite{pypi_registry}.
These markets are curated independently and differ in their listing policies, update frequency, and presentation formats, reflecting diverse community practices for publishing MCP servers.

Although these markets differ in presentation formats and available metadata, most share a common set of core information, such as project names, descriptions, and repository links.
We design customized crawlers for each market to extract these fields and retrieve the corresponding source repositories, enriching each entry with repository-level metadata.
Aggregating multiple sources mitigates the bias introduced by market-specific practices.
As the same MCP server may appear across markets, we de-duplicate entries based on normalized repository URLs for subsequent analysis.

However, during the collection process, we observed that certain repositories indexed by public markets are not actual MCP servers; instead, they may be empty or do not implement MCP interactions.
To ensure the quality of MCPZoo, we implement a lightweight validation step to remove projects that are not MCP servers.
As a fundamental requirement, a functional MCP server must support structured interaction through standardized transport mechanisms and comply with specific protocol handshakes. Consequently, we adopt a multi-stage filtering strategy.
First, we excluded documentation-only repositories and empty projects lacking executable source code. Next, we conducted a dependency analysis to confirm that the projects actively use MCP-related libraries. Specifically, we statically examine source code to verify the presence of server-side signatures, such as \texttt{@modelcontextprotocol/sdk/server} or \texttt{mcp.server.fastmcp}.
The process yields a refined set of candidate MCP servers, ready for large-scale deployment and analysis.

\subsection{Automated Deployment}

Due to varying developer expertise and maintenance quality, deploying MCP servers is challenging in practice.
At the ecosystem scale, manual diagnosis and repair are infeasible, motivating a systematic and automated solution.
To address this, we design a multi-agent build framework in which three specialized agents collaborate to iteratively construct MCP servers that can be reliably executed and analyzed under realistic runtime conditions.
These agents respectively generate deployment configurations, validate runtime behavior via protocol-level interactions, and analyze failures using execution feedback, enabling reliable and scalable analysis without manual intervention.
To ensure environment isolation, reproducibility, and reuse across deployments, we rely on Docker-based containerization to manage and preserve deployment artifacts.

\subsubsection{Dockerfile Generation}

The \textit{Generation Agent} aims to unify the mapping of diverse, non-standardized source code into uniform container deployment configurations. It is powered by a rich set of inputs, analyzing the project’s file tree, \texttt{README} documentation, configuration files, and the history of previous build errors. To ensure optimal build reliability, we designed a three-tiered pipeline.

\ttbf{Native Environment Reuse} Many developers include a Dockerfile to allow users to start the project with a single click. Therefore, our pipeline first checks whether a Dockerfile is already present in the project root. If so, it attempts to build and run the container without modification, in order to preserve the author's original execution environment. In practice, however, many Dockerfiles fail due to outdated dependencies or environment drift. Any failure at this stage automatically triggers a transition to the next stage.

\ttbf{Instruction Reproduce} When native deployment is unavailable, the agent switches to a lightweight generation strategy. Since installing published packages is inherently more stable than compiling raw source code, the agent analyzes the README for explicit configuration blocks or global installation instructions (e.g., \texttt{npm install}) and generates simplified Dockerfiles that install the application as a package.

\ttbf{Semantic Inference}
When global installation is not applicable, the agent builds the project directly from source code.
At a large scale, naive source-based builds are inefficient: they often introduce large build contexts and long build times, leading to substantial storage and network overhead.
To address this, we identify and solve three concrete challenges in the build process:

(1) \textit{Irrelevant file inclusion.}
Naive Copying of the entire repository often includes version control history (e.g., \texttt{.git}), local environments (e.g., \texttt{node\_modules}), or large test suites. These files increase the image size and
may introduce conflicts.
To solve this, the agent analyzes the file tree and documentation to understand the project structure,
and selectively includes only essential source files
This approach significantly reduces the build context size and prevents interference from local artifacts.

(2) \textit{Environment inconsistency.}
Non-standard or heavy base images frequently cause dependency conflicts and bloated builds.
The agent enforces a normalization strategy. It uses lightweight and standardized base images, e.g., \texttt{python:3.11-slim} or \texttt{node:18-alpine}. This solution optimizes build speed and minimizes storage usage. It also improves reliability by reducing the risk of version incompatibilities found in complex images.

(3) \textit{Entrypoint identification.}
Determining the correct startup command is non-trivial when relying on implicit conventions.
To ensure the application starts as the developer intended, the system employs a prioritized search. It first looks for explicit "Usage" or "Start" instructions in the \texttt{README} file,
and otherwise falls back to structured metadata, such as the \texttt{"start"} script in \texttt{package.json} or standard entry files. This validation ensures that the final container executes the correct application entry point.

\subsubsection{Protocol Verification}

The \textit{Verification Agent} is responsible for validating whether a server can support correct MCP interactions. It first builds and launches the server in a containerized environment, and then performs protocol-level verification. In addition to validation, the agent also captures structured failure signals to support downstream diagnosis and repair.

\ttbf{Protocol-Level Verification}
The agent instantiates a fully compliant \textit{MCP Client} to initiate a connection via the specific transport protocol (Stdio, SSE, or Streamable HTTP). The verification enforces a strict JSON-RPC 2.0 handshake sequence~\cite{anthropic2024mcp}
: establishing the transport connection, sending the \texttt{initialize} payload, and finally executing the \texttt{tools/list} capability check. The deployment is deemed successful only if the server responds with a valid, schema-compliant list of tools, thereby confirming that the MCP protocol is correctly implemented and operational.

\ttbf{Failure Capture}
To ensure process stability, the agent instantiates the container in an isolated sandbox and monitors its status for a post-startup period. This mechanism effectively detects immediate crash loops caused by runtime configuration issues, such as missing environment variables or permission denials, that typically evade static build-time checks. Crucially, if the process stops at any stage due to a build error or runtime crash, the agent automatically collects the logs and exit codes. This error data is then passed to the \textit{Diagnosis Agent} to trigger repairing.

\subsubsection{Failure Diagnosis}

Acting as the recovery engine, the \textit{Diagnosis Agent} transforms execution failures into targeted repair strategies. Rather than blindly retrying, this agent implements an evidence-based reasoning process, grounding every repair decision in specific error patterns to converge on a valid configuration.

\ttbf{Context Extraction}
Raw execution logs are often extremely long and contain a large amount of irrelevant information, which makes it difficult to locate the actual cause of a failure.
In practice, build processes typically terminate immediately upon facing a fatal error, causing the most informative error messages to appear near the end of the log.
Based on this observation, our goal is to isolate the specific error message that triggered the crash.
To achieve this, the agent applies a filtering mechanism to extract core segments, focusing on the tail of the build log and relevant stack traces.

\ttbf{Dependency Resolution}
A frequent cause of failure is missing software components in the container environment. Common errors include missing Python modules or unavailable system commands. The objective is to ensure the environment contains all necessary libraries and tools. To solve this, the agent employs a deterministic mapping mechanism. It translates specific failure signatures into targeted repair templates. For example, evidence of dependency gaps like \texttt{ModuleNotFound} Error triggers explicit installation directives such as \texttt{RUN pip install}. Similarly, \texttt{Command not found} errors prompt the injection of system-level installation templates using \texttt{apt-get}.

\ttbf{Version Stabilization}
Simply installing tools is not sufficient if versions conflict or commands use invalid syntax. Issues often arise from incompatible library versions or incorrect CLI arguments. To resolve these conflicts and enforce a stable execution state, the agent analyzes logs for version mismatches or invalid options. It then generates high-priority instructions to lock exact package versions or correct the command structure. These constraints are fed back to the \textit{Generation }, ensuring that the next attempt is guided by the specific diagnosis of the previous error.

\subsection{Implementation}

\subsubsection{Environment \& Configuration}

The MCPZoo framework is deployed on a cluster of 5 servers
(64-core CPUs, 256 GB memory each), where all build, verification, and diagnosis tasks are executed in the environments with a standardized Ubuntu 22.04 runtime for MCP servers.
To support automated deployment, the agents rely on an LLM, Qwen3-235B-A22B-Instruct, to generate deployment configurations, interpret failure signals, and guide iterative adjustments.
To improve robustness, each server is assigned a maximum of 5 build attempts, allowing the system to iteratively refine configurations and maximize deployment success at scale.

\subsubsection{Alignment on Interaction}

In practice, MCP servers support three distinct interaction mechanisms, i.e., stdio, SSE, and Streamable HTTP, which differ in initialization semantics, communication patterns, and lifecycle management.
To enable uniform downstream automation and testing, MCPZoo abstracts away these differences and exposes a unified interaction interface.
To achieve this, we implement a Kubernetes-based image and container management system that automatically starts, stops, and supervises MCP server containers.
On top of this system, we define a standardized control interface that maps unified commands to protocol-specific execution logic.
A FastAPI~\cite{fastapi}-based gateway layer bridges external control requests into containerized MCP servers, enabling remote interaction.
Special handling is required for stdio-based MCP servers, which rely on local pipe communication and do not natively support remote access.
To address this limitation, we employ Supergateway~\cite{supergateway} to forward stdio streams to designated network ports, effectively transforming local interactions into remotely accessible endpoints. This design allows stdio-based MCP servers to be integrated into the same remote interaction and batch testing workflow as SSE- and HTTP-based servers.

\subsubsection{Human Validation}

To complement the agent-based verification loop, we perform a lightweight human validation on a stratified random sample of 100 MCP server records, including 70 successful deployments and 30 deployment failures. Two researchers independently inspect the original repositories, generated deployment artifacts, execution logs, and protocol-level verification traces. For successful cases, they check whether the project is a valid MCP server and whether the trace contains a successful MCP initialization and \texttt{tools/list} response. They also verify successful end-to-end tool invocation on the sampled deployments by selecting non-destructive tools with locally satisfiable inputs. For failure cases, they check whether the diagnosed cause is supported by build, startup, or interaction logs. The researchers reach 98\% agreement (\(\kappa=0.95\)); disagreements are resolved through discussion. Overall, 100\% of sampled successful deployments are confirmed as correct, and 93.3\% of sampled failure decisions are consistent with the observed logs. The remaining two failure cases can be manually repaired with additional effort, suggesting that the automated pipeline is conservative rather than over-claiming deployability.
 
\section{Characterize MCP Ecosystem}

Building upon MCPZoo as a large-scale foundation that supports dynamic analysis of real-world MCP servers, we conduct an ecosystem-wide measurement of the MCP landscape. This section aims to characterize how MCP servers are developed, deployed, and used in practice, moving beyond repository-level statistics to capture properties that only display at runtime.

\begin{table}[t]
    \centering
        \caption{Scale of MCPZoo.}
    \label{tab:scale_and_source}
    \resizebox{\linewidth}{!}{
    \begin{tabular}{@{}lrrrr@{}}
    \toprule
    \textbf{Source} & \textbf{Raw} & \textbf{Valid} & \textbf{Dynamic} & \textbf{Dynamic \%} \\
    \midrule
    MCP Store & 39,770 & 28,960 & 14,870 & 51.35\% \\
    MCP World & 31,046 & 25,519 & 14,655 & 57.43\% \\
    MCP Market & 16,168 & 14,127 & 8,741 & 61.87\% \\
    MCP Repository & 14,341 & 12,263 & 8,066 & 65.78\% \\
    AI Base MCP & 11,120 & 9,275 & 6,154 & 66.35\% \\
    NPM & 18,560 & 8,295 & 4,759 & 57.37\% \\
    Pulse MCP & 6,885 & 6,248 & 4,415 & 70.66\% \\
    MCP.so & 6,772 & 5,241 & 3,429 & 65.43\% \\
    PyPI & 9,023 & 2,371 & 1,151 & 48.54\% \\
    Simthery & 3,157 & 1,628 & 967 & 59.40\% \\
    \midrule
    \textbf{Total} & \textbf{156,842} & \textbf{113,927} & \textbf{67,207} & \textbf{58.99\%} \\
    \textbf{Unique}  &
    -- &
    \textbf{64,611} &
    \textbf{37,288} & \textbf{57.71\%} \\
    \bottomrule
    \end{tabular}
    }
\end{table}

\subsection{Landscape}\label{sec:overlap}

As summarized in Table~\ref{tab:scale_and_source}, we collected 156,842 raw entries across ten major markets. After validation, we identified 113,927 valid MCP servers, of which 64,611 are unique. Notably, our automated framework successfully established interaction with 37,288 unique servers, achieving an overall interaction rate of 57.71\%.
Throughout the study, these 64,611 unique MCP servers are the main measurement set.
Detailed analysis of deployment failures is presented in Section~\ref{sec:deploy_result}.
These dynamic servers provide a solid foundation for understanding the functional maturity of the current ecosystem.

\begin{figure}[t]
\centering
\includegraphics[width=\linewidth]{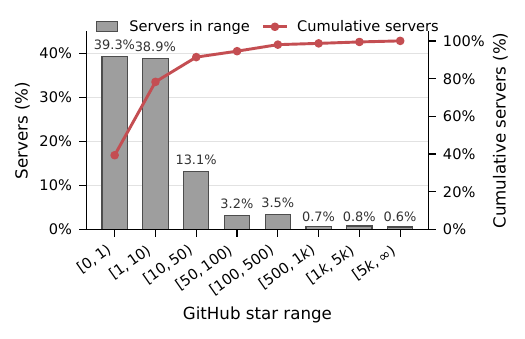}
\caption{GitHub star distribution of MCP server repositories.}
\Description{Bar chart and cumulative line showing a long-tailed GitHub star distribution for MCP server repositories.}
\label{fig:server_star_distribution}
\end{figure}

We further examine the popularity of MCP server repositories and find a clear long-tailed distribution (Figure~\ref{fig:server_star_distribution}).
Most servers receive limited attention: 78.26\% have fewer than 10 stars.
Meanwhile, the upper tail still contains 386 servers (0.56\%) with at least 5,000 stars, with the most-starred repository, \texttt{n8n-io/n8n}, reaching over 195k stars.
This confirms that the MCP ecosystem contains a small number of highly visible servers and a large population of individually maintained or lightly adopted projects.

\begin{figure}[t]
\includegraphics[width=0.9\linewidth]
{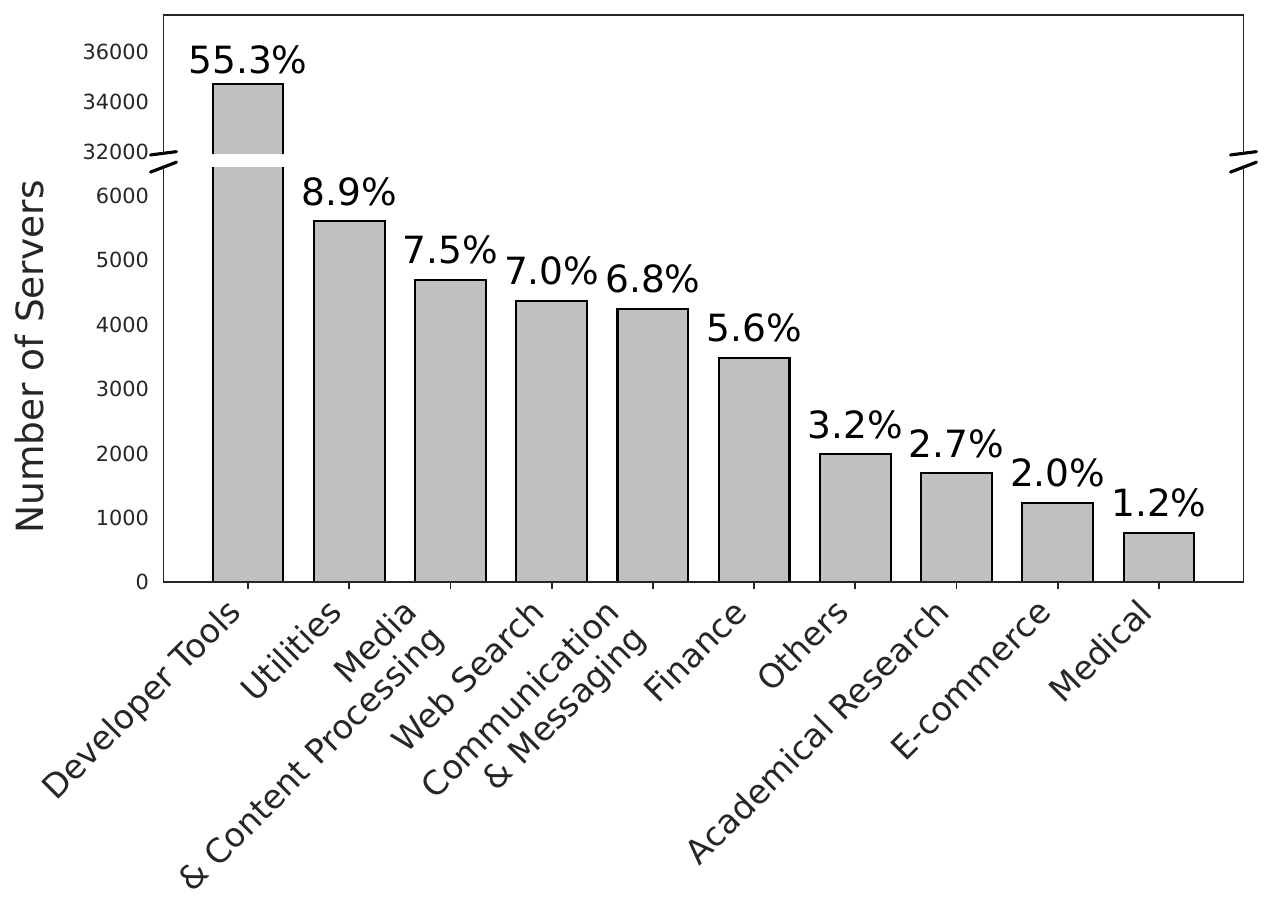}
\caption{MCP Servers across Different Functional Domains.}
\Description{Chart showing the distribution of MCP servers across different functional domains.}
\label{fig:server_func_count}
\end{figure}

\ttbf{Server Function}\label{sec:server_func}
To characterize the diversity of application scenarios in the MCP ecosystem,
we classified MCP servers into nine primary functional domains (e.g., \textit{Finance}, \textit{E-commerce}, and \textit{Developer Tools}), based on their primary usage.
Our taxonomy follows prior MCP study~\cite{ray2025survey}.
The classification is performed using an LLM and validated via manual inspection (Appendix~\ref{appendix:evaluation-prompts}).
As illustrated in Figure~\ref{fig:server_func_count},
\textit{Developer Tools} dominate the ecosystem, accounting for 53.69\% of all servers, reflecting the protocol's primary adoption among technical users for coding assistance.
Although \textit{Finance} (5.40\%) and \textit{E-commerce} (1.91\%) servers are fewer, their direct interaction with monetary systems and transaction logic implies that successful attacks could cause immediate and significant economic losses. Similarly, \textit{Medical} servers (1.18\%) present critical safety risks, where compromise could lead to misleading medical advice, resulting in severe physical health consequences. Similarly, the widespread prevalence of \textit{Developer Tools} introduces systemic risks, as their misuse can facilitate improper database operations, backdoor insertions, or malicious code propagation.

\begin{table*}[t]
\centering
\caption{Distribution of Deployment Outcomes Based on Artifact Availability. }
\label{tab:deployment-overview}

\begin{tabular}{ccc| cc cc cc |cc}
\toprule
\multicolumn{3}{c|}{\textbf{Artifacts}} & \multicolumn{2}{c}{\textbf{Native Success}} & \multicolumn{2}{c}{\textbf{Agent Success}} & \multicolumn{2}{c}{\textbf{Failure}} & \multicolumn{2}{|c}{\textbf{Overall}} \\
\cmidrule(lr){1-3} \cmidrule(lr){4-5} \cmidrule(lr){6-7} \cmidrule(lr){8-9} \cmidrule(lr){10-11}
\textbf{Docker} & \textbf{Readme} & \textbf{Config} & \textbf{\#} & \textbf{\%} & \textbf{\#} & \textbf{\%} & \textbf{\#} & \textbf{\%} & \textbf{\#} & \textbf{\%} \\
\midrule

$\bullet$ & $\bullet$ & $\bullet$ &
1,769 & 19.6\% & 4,491 & 49.9\% & 2,747 & 30.5\% & 9,007 & 13.9\% \\

$\bullet$ & $\bullet$ & $\circ$ &
647 & 13.7\% & 1,733 & 36.6\% & 2,352 & 49.7\% & 4,732 & 7.3\% \\

$\bullet$ & $\circ$ & $\circ$ &
12 & 10.5\% & 46 & 40.4\% & 56 & 49.1\% & 114 & 0.2\% \\

$\circ$ & $\bullet$ & $\bullet$ &
-- & -- & 17,857 & 64.4\% & 9,850 & 35.6\% & 27,707 & 42.9\% \\

$\circ$ & $\bullet$ & $\circ$ &
-- & -- & 10,150 & 46.8\% & 11,556 & 53.2\% & 21,706 & 33.6\% \\

$\circ$ & $\circ$ & $\circ$ &
-- & -- & 583 & 43.3\% & 762 & 56.7\% & 1,345 & 2.1\% \\

\midrule
\multicolumn{3}{c|}{\textbf{Total}} & \textbf{2,428} & \textbf{3.8\%} & \textbf{34,860} & \textbf{54.0\%} & \textbf{27,323} & \textbf{42.3\%} & \textbf{64,611} & \textbf{100.0\%} \\
\bottomrule
\end{tabular}
\end{table*}

\begin{figure}[t]
    \centering
    \includegraphics[width=0.95\linewidth]{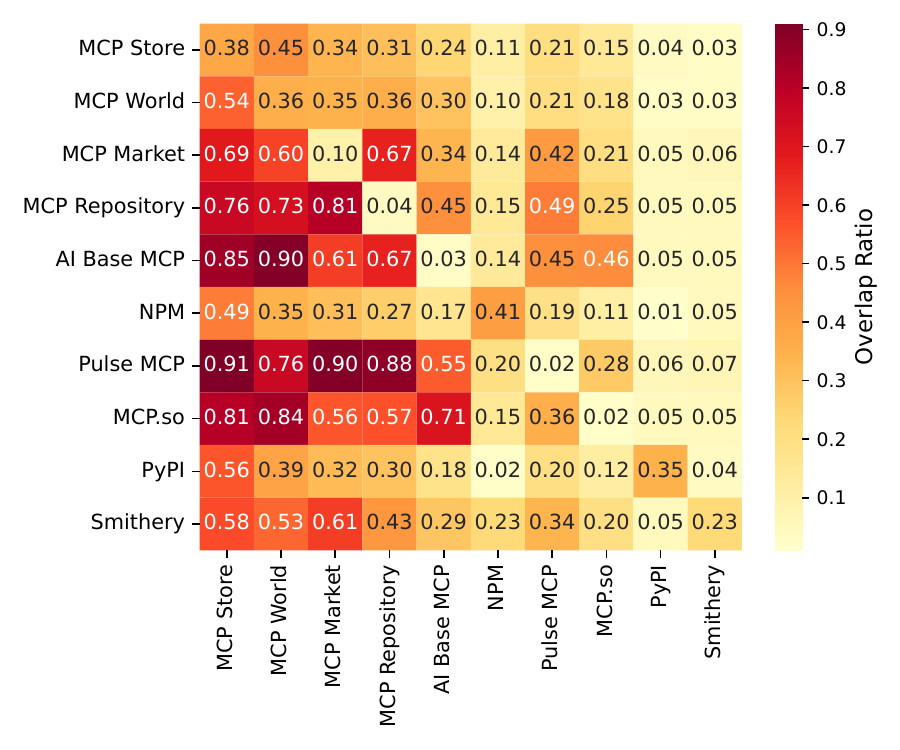}
    \caption{Intersection Ratio Matrix of MCP Servers across 10 Markets. Each cell $(i, j)$ represents the proportion of servers in Market $i$ (row) that are also indexed by Market $j$ (column). Diagonal cells $(i,i)$ indicate the proportion of servers exclusive to Market $i$. }
    \Description{Heatmap showing pairwise intersection ratios of MCP servers across 10 markets.}
    \label{fig:overlap_ratio}
\end{figure}

\ttbf{Cross-Market Overlap}
Although each market claims to offer a rich collection of MCP servers,
it is unclear whether their offerings provide unique or representative coverage of the ecosystem.
We calculated the overlap between sources to characterize cross-market overlap, as illustrated in Figure~\ref{fig:overlap_ratio}.
Most repositories exhibit severe overlap; notably, PulseMCP shows the highest, with 91\% of its servers also appearing on MCP Store, suggesting strong redundancy between markets. Conversely, the overlap between PyPI and NPM is minimal, sharing only 32 common servers (1\%).
This overlap corresponds to projects that provide both Python and TypeScript/JavaScript implementations, indicating that developers typically choose distribution markets based on the target implementation language.
To further assess market uniqueness, we calculated the Market Exclusivity Rate ($E_r$) for each market $P$:
\begin{equation}
E_r(P) = \frac{| \{ e \in S_P \mid \forall P' \neq P, e \notin S_{P'} \} |}{| S_P |}
\end{equation}
As shown by the diagonal entries in Figure~\ref{fig:overlap_ratio},
NPM maintains the highest proportion of new projects, while MCP Store (38\%) and MCP World (36\%) serve as the primary hubs for original emerging MCP servers among general aggregators.
In contrast, although MCP Market hosts a large number of servers, it exhibits substantial overlap with other markets, with only 10\% of servers excluded.

\finding{
Up to 90\% of servers in some MCP markets are shared across markets, indicating severe cross-market overlap.
}

\ttbf{Duplication}
While cross-market overlap reflects redundancy across markets, duplication further reveals repeated reuse of the same code in development.
In practice, MCP servers are often forked, copied, or re-published through mirrors with minimal modifications.
To better capture the true scale of the ecosystem, we focus on code-level duplication beyond the URL-based deduplication described in Section~\ref{sec:server_collection}.
Specifically, we consider three common duplication patterns observed in the wild.
First, for forked repositories, we trace their origin using the GitHub REST API~\cite{github_rest_api} and map them to the corresponding upstream repository.
Second, for direct code reuse and mirror repositories, which often do not preserve explicit fork relationships, we extract the original repository reference links from README files or project descriptions when available.
By resolving these relationships, we group MCP servers that share the same underlying codebase into a single entity, enabling a more accurate measurement of ecosystem scale and duplication.

The result reveals widespread duplication in the MCP ecosystem.
Among the 64,611 collected MCP servers, we identify only 46,327 unique entities, meaning that 28.3\% of unique servers are duplicate representations caused by forks, copies, or mirrors.
Notably, 3,973 entities appear multiple times within the same market, indicating that duplication is not solely driven by cross-market re-publication.
A notable pattern behind duplication is
template-driven development.
We observe large clusters of nearly unmodified replicas derived from shared templates.
The largest cluster involves Cloudflare Worker templates, particularly \texttt{remote-mcp-server-authless}, which accounts for 537 identical instances,
including 460 replicas within mcpworld.com.
Surprisingly, there are seven templates that each produce more than 50 replicas within a single market.
These findings indicate that a significant portion of the ecosystem's growth is fueled by low-effort replication rather than original innovation.

\finding{
28.3\% of unique MCP servers are duplicate representations caused by forks, copies, or mirrors.
}

\subsection{Deployment Feasibility}\label{sec:deploy_result}

Understanding the security implications of MCP servers requires that their behavior can be reliably observed and validated under realistic execution conditions.
However, in practice, many MCP server projects cannot be deployed or reproduced without substantial effort, often due to code quality issues.
We therefore examine the reproducibility and deployment characteristics of MCP servers in the ecosystem, focusing on how common engineering practices affect the feasibility of dynamic analysis.

We perform a source-level analysis of all 64,611 collected MCP server projects, examining the presence of deployment artifacts and configuration metadata.
As summarized in Table~\ref{tab:deployment-overview}, 78.6\% of the projects lack an explicit \textit{Dockerfile}, suggesting that containerization is not yet a standard practice within the MCP community. More importantly, a given Dockerfile does not guarantee runnability. Even for projects with a complete set of files (Dockerfile, README, and Configuration), the native success rate remains low at 19.6\%, highlighting a widespread lack of ``out-of-the-box'' reproducibility.

In contrast, our framework significantly bridges this gap, achieving an overall \textit{Agent Success} rate of 54.0\%. We found that structured configuration is the most critical factor for recovery.
Surprisingly, repositories lacking a Dockerfile but providing a Config achieved a 64.4\% success rate, which is higher than the 49.9\% rate of projects equipped with existing Dockerfiles. This suggests that existing Dockerfiles can be misleading. Since many are outdated, the agent often wastes effort trying to fix broken code. In contrast, when no Dockerfile is present, the agent builds a clean, standard environment from scratch using the reliable configuration data.
Notably, even projects missing all three artifacts (Dockerfile, README, and Config) still achieved a meaningful success rate through agent-driven inference. Overall, these results demonstrate that structured configuration is a more reliable predictor of deployment success than container scripts or textual documentation alone.

Deployment failures arise from multiple factors, among which external dependencies are the dominant cause, accounting for 63.6\% of all failures.
These dependencies mainly fall into two categories.
First, some servers require API credentials to function.
For instance, the \textit{gsuite}~\cite{mcp_gsuite} server necessitates valid OAuth tokens, revealing that these projects are not standalone utilities but are tightly coupled with external proprietary services.
Second, other servers depend on pre-existing infrastructure or system environments.
A clear example is \textit{mcp-nomad}\cite{mcp_nomad}, which functions as a controller for HashiCorp Nomad. Such servers presuppose access to a local configuration file or a live connection to an active cluster.
These prerequisites make them inherently verifying-resistant within a hermetic sandbox environment, as they require an operational infrastructure context to initialize.
In addition, 15.7\% of failures are caused by original source code defects or misconfigurations unrelated to the environment.

\finding{
78.6\% of MCP server projects lack Dockerfiles, and even among projects with complete deployment artifacts (Dockerfile, README, and Config), native deployment succeeds for only 19.6\%, indicating low out-of-the-box reproducibility across the ecosystem.
}

\begin{table*}[t]
    \centering
        \caption{Distribution of MCP Tool Capabilities.}
    \label{tab:mcp_tool_risks}
    \begin{tabular}{@{}c l p{7.5cm} r r@{}}
    \toprule
    \textbf{Risk Level} & \textbf{Capability Category} & \textbf{Tool Examples} & \textbf{\#} & \textbf{\%} \\
    \midrule

        \multirow{3}{*}{\textbf{High}}
      & Command \& Script Execution & Shell commands, Scripts (JS/Python), Database queries & 24,866 & 6.72\% \\
      & Outbound Data Transfer & File upload, Email/SMS dispatch, Social posting & 77,872 & 21.03\% \\
      & Local File Modification & File create/delete/modify, Permission change & 36,676 & 9.91\% \\
    \cmidrule{1-5}
        \multirow{3}{*}{\textbf{Medium}}
      & Local Info Retrieval & File reading, Screen capture, Audio recording & 28,737 & 7.76\% \\
      & Remote Info Retrieval & Web browsing, Remote fetch & 174,771 & 47.21\% \\
      & Prompt Providing & Providing popular system prompts for LLM & 3,873 & 1.05\% \\
    \cmidrule{1-5}

        \textbf{Low} & Others & Math evaluation, Data encoding & 23,424 & 6.33\% \\

    \bottomrule
    \end{tabular}
\end{table*}

\subsection{Tool}\label{sec:tool}

Tools serve as the fundamental execution units of the MCP ecosystem, turning static servers into dynamic agents capable of interacting with external environments. Since these tools strictly define the boundaries of permissible actions, the exposed tools directly shape the system's functional surface and attack surface.

\ttbf{Tool Scale}
We analyze the number of tools per server and observe a clear long-tail distribution.
85.53\% of servers offer fewer than 10 tools,
focusing on specific, narrow tasks.
In contrast, a small number of "super-servers" expose a very large number of tools.
For example, \textit{mcp-allthetools} aggregates 1,061 projects sourced from glama.ai, culminating in a total of 8,816 tools.
Other high-volume servers include \textit{amazon-elastic-compute-cloud} (1,182 tools), \textit{loketnl-api} (641 tools), and \textit{Meraki Dashboard API} (615 tools), which expose extensive interfaces and act as comprehensive gateways to external services with a large number of APIs.

\ttbf{Tool Capability}
Since the MCP server behavior is fundamentally determined by the tools they expose,
capability-level categorization provides a direct lens for assessing both functionality and potential abuse.
We categorize tools into seven capability groups following prior work~\cite{guo2025systematicanalysismcpsecurity}, based on interaction medium and data flow, and further organize them into three risk levels, enabling systematic measurement of capability distribution and associated risks.
Details of classification and validation are provided in Appendix~\ref{appendix:evaluation-prompts}.
As shown in Table~\ref{tab:mcp_tool_risks}, High-risk tools account for 37.66\% of all capabilities, while Medium-risk tools dominate with 56.02\%, indicating a widespread presence of potentially risky functionality in the ecosystem.
High-risk categories, such as Command \& Script Execution, Local File Modification, and Outbound Data Transfer, enable direct system manipulation or data exfiltration. These capabilities can directly alter system state or leak sensitive information.
Medium-risk capabilities include Local Info Retrieval, Remote Info Retrieval, and Prompt Providing.
These capabilities primarily expose sensitive data or introduce indirect attack surfaces, such as prompt injection.

\begin{figure}[t]
\includegraphics[width=\linewidth]{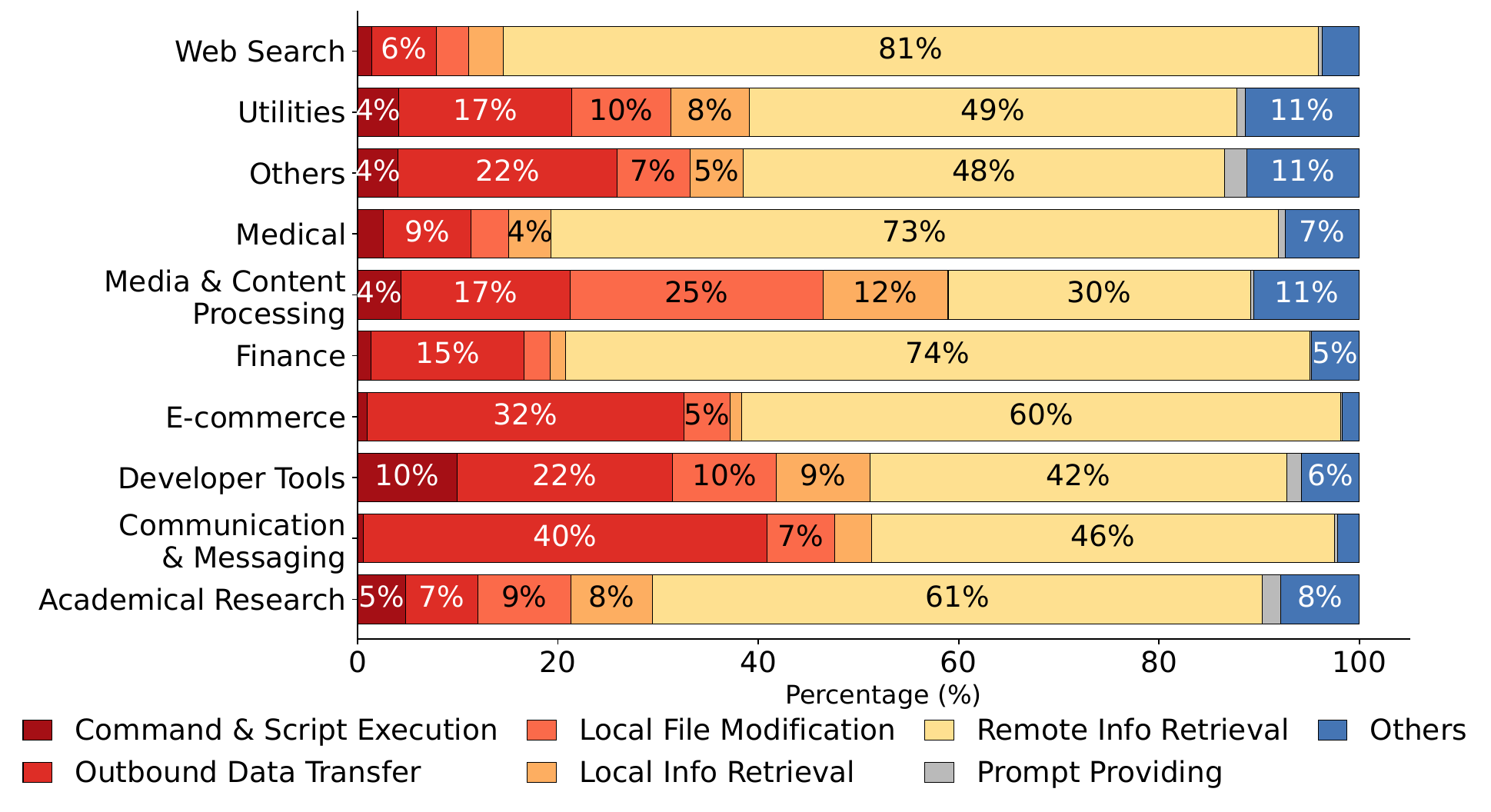}
\caption{Tool Capabilities within Each Server Function.}
\Description{Chart showing how MCP tool capabilities are distributed within each server function.}
\label{fig:capa_func}
\end{figure}

\ttbf{Server Function vs Tool Capability} We also investigated the relationship between server function and tool capability, revealing that distinct functions exhibit unique capability distributions driven by operational requirements (Figure~\ref{fig:capa_func}). Servers in \textit{Web Search} (81.37\%), \textit{Finance} (74.34\%), \textit{Medical} (72.56\%), and \textit{Academical Research} (60.87\%) categories contain a dominant proportion of Remote Info Retrieval capabilities to fetch real-time data, rendering them particularly susceptible to indirect prompt injection attacks.  \textit{Communication \& Messaging} and \textit{E-commerce} servers allocate a significant percentage of their tools to Outbound Data Transfer (40.31\% and 31.61\% respectively) to manage communications and orders, thereby becoming attractive targets for attackers aiming to exfiltrate sensitive user data. \textit{Media \& Content Processing} servers possess a comparatively high density of tools with Local File Modification capabilities (25.26\%), a surface that attackers can exploit to tamper with local file systems or deploy ransomware. Finally, \textit{Developer Tools} account for 83.15\% of the ecosystem's total Command \& Script Execution capabilities; while necessary for deployment and testing, this concentration also creates a critical vector for attackers.

\finding{ Distinct server functions exhibit unique tool capability distributions that create domain-specific risks. Specifically, Web Search servers rely heavily on retrieval tools 81.37\%, while Developer Tools concentrate on command execution capabilities 9.92\%. }

\section{Reliability of MCP Security Scanners}

\begin{table*}[t]
\centering
\caption{Scanner-Reported Risk Statistics across Different MCP Scanners.}
\label{tab:vulnerability_results_dedup_server_url}
\resizebox{\textwidth}{!}{
\begin{tabular}{l cccc ccc}
\toprule
\multirow{2.5}{*}{\textbf{Scanner}}
& \multicolumn{4}{c}{\textbf{Reported Categories}}
& \multirow{2.5}{*}{\makecell{\textbf{Reported}\\ \textbf{Risk}}} & \multirow{2.5}{*}{\makecell{\textbf{Analyzed}\\ \textbf{Servers}}} & \multirow{2.5}{*}{\makecell{\textbf{Reported}\\ \textbf{Risk \%}}} \\
\cmidrule(lr){2-5}
& \textbf{Prompt Injection}
& \textbf{Command Execution}
& \textbf{Data Leakage}
& \textbf{Others}
& & & \\
\midrule
Agent-Scan~\cite{mcp_scan2025} & 0.72\% & 30.86\% & 19.74\% & 23.73\% & 15,199 & 33,107 & 45.91\% \\
A.I.G (static)~\cite{Tencent_AI-Infra-Guard_2025} & 5.28\% & 18.89\% & 10.79\% & 16.36\% & 15,957 & 36,827 & 43.33\% \\
A.I.G (dynamic) & 39.58\% & 6.26\% & 5.35\% & 53.37\% & 20,761 & 34,017 & 61.03\% \\
MCP-Scanner~\cite{mcp-scanner-cisco} & 22.20\% & 3.14\% & 7.75\% & -- & 8,805 & 33,218 & 26.51\% \\
MCPScan~\cite{mcpscan-antgroup} & 76.58\% & 31.92\% & 1.85\% & -- & 28,220 & 35,259 & 80.04\% \\
MCPSafetyScanner~\cite{mcpsafetyscanner} & 0.08\% & 0.19\% & 0.24\% & 0.54\% & 176 & 32,585 & 0.54\% \\
mcp-gateway~\cite{mcp-gateway} & 0.02\% & 1.38\% & 1.27\% & -- & 971 & 37,288 & 2.60\% \\
nova-proximity~\cite{nova-proximity} & 8.79\% & -- & -- & -- & 3,210 & 36,501 & 8.79\% \\
mcp-armor~\cite{mcp-armor} & 39.21\% & 5.48\% & -- & -- & 15,040 & 37,082 & 40.56\% \\
\bottomrule
\end{tabular}
}
\begin{minipage}{0.95\linewidth}
\footnotesize{
\textbf{Note:} ``--'' indicates this scanner does not report this category. Reported risks are scanner outputs and should not be interpreted as validated vulnerabilities. \\
}
\end{minipage}
\end{table*}

This section evaluates whether existing MCP security scanners can reliably support ecosystem-scale security conclusions.
We run eight representative scanners on the runtime-enabled MCPZoo dataset, compare their reported risks, and then validate whether those reports correspond to real vulnerabilities.
Our goal is not to treat scanner outputs as ground truth, but to measure how scanner design choices, execution settings, and input assumptions affect the reliability of MCP security reports.

\subsection{Experiment Setup}

\ttbf{Scanner Selection}
To the reliability of existing MCP security scanners, we select eight representative scanners that are publicly available and widely used in practice.
Concretely, we include scanners with more than 200 GitHub stars to ensure basic maturity and reproducibility, while also ensuring coverage across different analysis paradigms, including static analysis, dynamic interaction, and hybrid approaches.
Specifically, we evaluate Agent-Scan~\cite{mcp_scan2025}, A.I.G~\cite{Tencent_AI-Infra-Guard_2025}, MCP-Scanner~\cite{mcp-scanner-cisco}, MCPScan~\cite{mcpscan-antgroup}, MCPSafetyScanner~\cite{mcpsafetyscanner}, mcp-gateway~\cite{mcp-gateway}, nova-proximity~\cite{nova-proximity}, and mcp-armor~\cite{mcp-armor}.
These scanners cover different analysis strategies.
MCPScan performs purely static analysis on source code. The other scanners require a live MCP server and perform dynamic analysis by interacting with it via the MCP protocol. Additionally, A.I.G supports both static and dynamic modes.
However, these dynamic scanners differ in interaction depth, ranging from simple metadata retrieval (e.g., tool listing) to more active probing through tool invocation.
Detailed comparisons are provided in Appendix~\ref{sec:scanner_logic}.

\ttbf{Scanner Execution Workflow}
We focus on the 37,288 interactable MCP servers in MCPZoo, as dynamic scanners require a valid runtime interface and cannot be applied to non-interactable servers.
For each server, we prepare two standardized inputs: the source code and a runtime endpoint exposed through MCPZoo's unified interaction layer.
To ensure fairness and reproducibility,  all scanners are executed under a controlled environment,
with standardized inputs, identical runtime constraints, and consistent resource limits, using default configurations without task-specific tuning.
For scanners relying on an LLM, we use a fixed local Qwen3-235B-A22B-Instruct deployment to eliminate variability from different model backends.
We set a maximum execution timeout of 1500 seconds and limit each scanner to at most 5 iterations per server, balancing coverage and efficiency.
Detailed settings for each scanner are summarized in Table~\ref{tab:scanner_setup}.

\subsection{Result Analysis}\label{sec:result_analysis}

\ttbf{Overall Scanner Reports}
Table~\ref{tab:vulnerability_results_dedup_server_url} reports the eight scanner outputs across all eight scanners.
Across all scanners, 96.89\% of the 37,288 interactable MCP servers are reported as risky by at least one scanner, yet strikingly, none of these servers are flagged by every single one of the eight scanners.
This large gap indicates that scanner reports are highly sensitive to scanner design and should not be interpreted as direct evidence that those servers are truly vulnerable.

\begin{table}[t]
\centering
\caption{Scanner-reported risk rates by GitHub star range.}
\label{tab:top_star_risk}
\resizebox{\linewidth}{!}{
\begin{tabular}{lrrr}
\toprule
GitHub Stars & Scanned Servers & Risky Servers & Reported Risk \% \\
\midrule
{[5,000,$\infty$)} & 171 & 167 & 97.66\% \\
{[1,000,5,000)} & 258 & 252 & 97.67\% \\
{[500,1,000)} & 226 & 222 & 98.23\% \\
{[100,500)} & 1,343 & 1,325 & 98.66\% \\
{[50,100)} & 1,329 & 1,303 & 98.04\% \\
{[10,50)} & 5,453 & 5,336 & 97.85\% \\
{[1,10)} & 14,186 & 13,736 & 96.83\% \\
{[0,1)} & 14,322 & 13,788 & 96.27\% \\
\midrule
All scanned & 37,288 & 36,129 & 96.89\% \\
\bottomrule
\end{tabular}
}
\begin{minipage}{0.95\linewidth}
\footnotesize{\textbf{Note:} Reported risk denotes at least one scanner flag among scanned servers.}
\end{minipage}
\end{table}

We further group scanned servers by GitHub stars to examine whether more popular repositories appear safer.
As shown in Table~\ref{tab:top_star_risk}, reported risk rates remain similarly high across star ranges, from 96.27\% in the [0,1) range to 97.66\% among servers with at least 5,000 stars.
This suggests that higher repository popularity does not correspond to a substantially lower scanner-flag rate.

For scanner coverage and execution reliability, static scanners can analyze most servers, except when large repositories exceed model context limits.
Additionally, MCPScan is sensitive to non-UTF-8 encoded content, which can cause processing failures.
In contrast, dynamic scanners exhibit lower coverage. Agent-Scan, for instance, depends on vendor-hosted backends that may return HTTP 429 (Too Many Requests) errors. To avoid overloading these services, we limit request rates.
More generally, dynamic scanners often fail due to timeouts, protocol incompatibility, missing credentials (e.g., API keys), or external service rate limits.

To enable systematic comparison across scanners, we harmonize their different output schemas into four unified risk categories: \textit{Prompt Injection}, \textit{Command Execution}, \textit{Data Leakage}, and \textit{Other}. The mapping is summarized in Appendix \ref{appendix:risk-category-mapping}. \textit{Prompt Injection} refers to attacks in which adversarial instructions are embedded into any content that enters the LLM's context window with the goal of hijacking the agent's control flow and inducing unintended behavior. \textit{Command Execution} captures risks where an agent is manipulated into executing arbitrary system commands, shell scripts, or dynamically generated code, potentially causing irreversible damage to the underlying system. \textit{Data Leakage} covers the unauthorized disclosure of sensitive information such as API keys, tokens, and passwords embedded in tool source code or configuration files.

We next analyze the distribution of reported risks across scanners. A striking observation is the large variation in reported risk rates, ranging from 0.54\% (MCPSafetyScanner) to 80.04\% (MCPScan), indicating substantial divergence in scanner outputs.
This inconsistency is particularly pronounced for \textit{Prompt Injection}, whose reported rates vary from 0.02\% (mcp-gateway) to 76.58\% (MCPScan). Such variation suggests that the detection of prompt injection is highly dependent on scanner-specific strategies.
Moreover, scanners differ not only in detection rates but also in coverage: some scanners do not report certain categories at all, indicating differences in analysis scope.
Overall, these results show that scanner outputs vary significantly in terms of sensitivity, coverage, and reporting, raising concerns about the consistency and reliability of reported risks, which we validate in Section~\ref{sec:validation}.

\finding{
Although existing scanners flag 96.89\% of MCP servers as potentially risky, their conclusions differ widely in both server coverage and reported categories.
}

\begin{figure}[t]
\centering
\includegraphics[width=\linewidth]{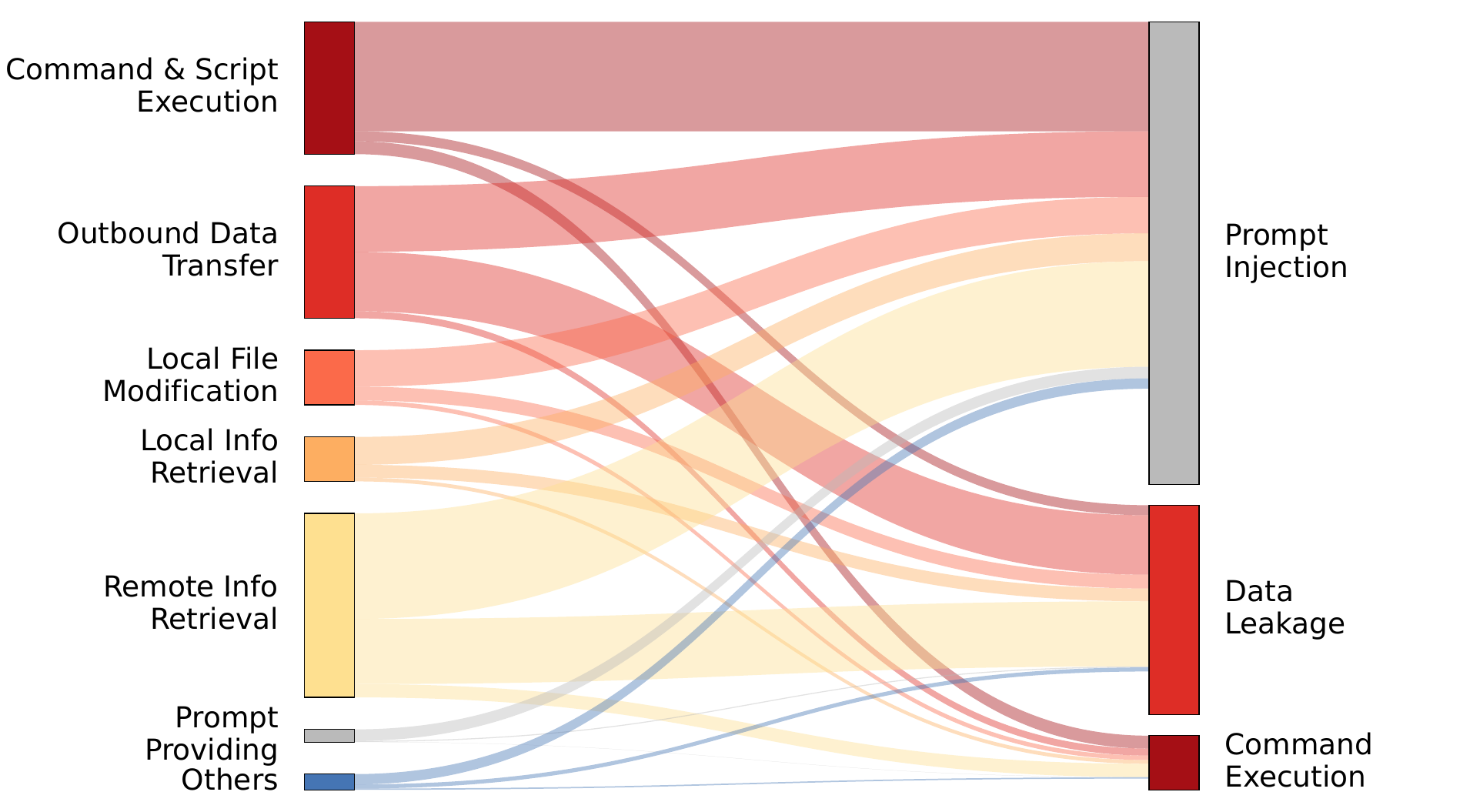}
\caption{The Relationship between MCP Tool Capabilities (Left) and Scanner-Reported Risk Categories (Right).}
\Description{Sankey-style visualization linking MCP tool capabilities to scanner-reported risk categories.}
\label{fig:tool_capability_risk}
\end{figure}
\begin{figure}[t]
\centering
\includegraphics[width=\linewidth]{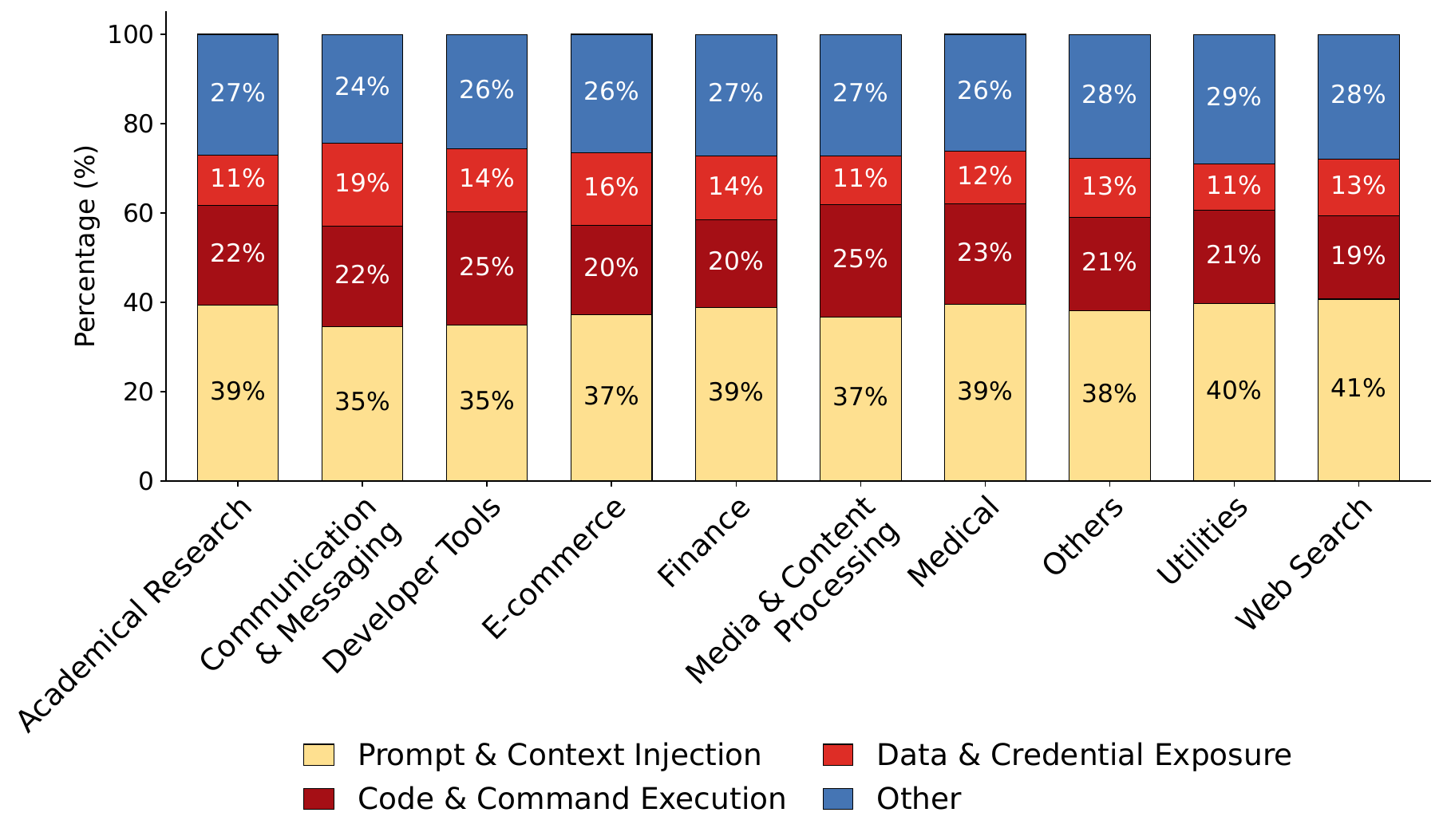}
\caption{Scanner-Reported Risks within Different Types of MCP Server Functions.}
\Description{Chart showing scanner-reported risks across different types of MCP server functions.}
\label{fig:func_risk}
\end{figure}

\ttbf{Tool Capability \& Server Function}
To characterize associations in scanner reports, we analyze the relationships among MCP tool capabilities~(in Section~\ref{sec:tool}), MCP server functions~(in Section~\ref{sec:server_func}), and scanner-reported risk categories.

At the capability level, reported risks exhibit clear structural patterns. As shown in Figure~\ref{fig:tool_capability_risk}, different tool capabilities are consistently associated with distinct risk categories. For example, network data reading and sending contribute 74.91\% of Data Leakage risks, while 82.72\% of execution-related capabilities are linked to \textit{Prompt Injection} reports. This suggests that certain capability classes are consistently associated with specific types of reported risks.

In contrast, the distribution of risks across server functions is relatively uniform. As shown in Figure~\ref{fig:func_risk}, different application domains exhibit similar proportions of risk categories, with only moderate variation across domains.
These patterns show that scanner-reported risks vary more clearly across tool capabilities than across high-level application domains.

\finding{
Scanner-reported risks exhibit capability-associated patterns and vary comparatively little across application functions.
}

\subsection{Validation of Scanner Reports}\label{sec:validation}

The preceding analysis suggests that scanner outputs may not reliably reflect underlying risks. We therefore conduct a validation study to assess their precision, recall, and cross-scanner consistency.

\ttbf{Precision Validation}
First, we randomly sample 100 MCP servers that are successfully processed by scanners and reported as risky by at least one scanner.
The sample is stratified by scanner and risk category to reflect per-scanner report distributions while covering all categories, so that high-volume scanners or categories do not dominate the manual review.
For each sampled server, two reviewers independently inspect the scanner evidence, source code, runtime metadata, and when possible, execute the relevant MCP interaction in MCPZoo.
A report is labeled as a true positive only when the evidence demonstrates a concrete vulnerable behavior or a reachable unsafe data or control flow. Reports based only on broad capability, suspicious wording, or unverifiable model inference are labeled as false positives or unverifiable.
The initial reviewer agreement is substantial (Cohen's $\kappa=0.87$), and disagreements are resolved by a third reviewer.

As shown in Table~\ref{tab:scanner_validation},
overall average precision is 45.53\%, with substantial variation across scanners (10.40\%--96.88\%).
We observe
that scanners reporting only a small number of risky servers tend to achieve substantially higher precision.
In particular, MCPSafetyScanner, mcp-gateway, and nova-proximity flag far fewer servers than the other scanners, yet consistently produce more precise alerts. By contrast, scanners that report risks at a larger scale generally exhibit noticeably lower precision,
an observation consistent with a possible tradeoff between coverage and accuracy in our sample.
We also observe
differences across analysis strategies. Code-based scanners (e.g., MCPScan and A.I.G (static)) tend to achieve higher precision than interaction-driven scanners (e.g., Agent-Scan and A.I.G (dynamic)).

\begin{table}[t]
\centering
\caption{Precision and Recall of Scanner-Reported Risks. }
\label{tab:scanner_validation}
\resizebox{\linewidth}{!}{
\begin{tabular}{l c c c c c c}
\toprule
\multirow{2}{*}{\textbf{Scanner}} & \multicolumn{3}{c}{\textbf{Precision Validation}} & \multicolumn{3}{c}{\textbf{Recall Validation}} \\
\cmidrule(lr){2-4}\cmidrule(lr){5-7}
& \textbf{Sampled} & \textbf{TP} & \textbf{Precision} & \textbf{Affected} & \textbf{TP} & \textbf{Recall} \\
\midrule
Agent-Scan~\cite{mcp_scan2025}         & 78  & 22 & 28.21\%   & 32 &  16 &  50.00\% \\
A.I.G (static)~\cite{Tencent_AI-Infra-Guard_2025}  &  86 & 49 & 56.98\% & 34 &  16 &  47.06\% \\
A.I.G (dynamic)~\cite{Tencent_AI-Infra-Guard_2025} & 125 & 13 & 10.40\% & 34 &  2 &  5.88\% \\
MCP-Scanner~\cite{mcp-scanner-cisco}   &  69 & 17 & 24.64\% & 33 &  7 &  21.21\% \\
MCPScan~\cite{mcpscan-antgroup}        & 123 & 56 & 45.53\% & 35 &  26 & 74.29\% \\
MCPSafetyScanner~\cite{mcpsafetyscanner} & 77 & 57 & 74.03\%  & 26 &  0 &  0.00\% \\
mcp-gateway~\cite{mcp-gateway}         & 32  & 31 & 96.88\%   & 35 &  0 &  0.00\% \\
nova-proximity~\cite{nova-proximity}   & 36  & 19 & 52.78\%   &  4 &  0 &  0.00\% \\
mcp-armor~\cite{mcp-armor}             & 59  & 12 & 20.34\%   & 21 & 4 & 19.05\% \\
\midrule
\textbf{Overall} &  &  & 45.53\% &  &  & 24.17\% \\

\bottomrule
\end{tabular}
}
\begin{minipage}{0.98\linewidth}
\footnotesize{\textbf{Note:} For precision validation, ``Sampled'' counts alerts (not servers) from 100 sampled servers, and may exceed 100.}
\end{minipage}
\end{table}

\ttbf{CVE-Based Recall Validation}
To obtain a set of real-world vulnerable MCP servers, we construct a ground-truth dataset based on publicly disclosed CVEs from the NVD database.
We identify relevant CVEs using keywords such as \textit{Model Context Protocol} and map them to MCPZoo servers through exact repository matching and path-level alignment.
To determine vulnerability status, we compare each server’s download time with the corresponding fix time derived from GitHub releases, tags, or patch commits, labeling pre-fix instances as likely vulnerable.
After filtering to retain vulnerability types within our scope, this process yields a high-confidence dataset consisting of 10 CVEs affecting 38 MCPZoo servers (Appendix~\ref{sec:cve_list}).
Although not exhaustive, this dataset captures confirmed real-world vulnerabilities under current MCP practices and serves as a reliable reference for recall evaluation.

As shown in Table~\ref{tab:scanner_validation}, recall varies significantly across scanners. MCPScan identifies 74.29\% of known vulnerable servers,
while several scanners do not detect any cases in this benchmark.
In particular, mcp-gateway, nova-proximity, and MCPSafetyScanner, despite their relatively high precision, do not identify any vulnerable servers.
This contrast highlights a clear precision and recall tradeoff: scanners that report fewer, cleaner alerts can appear highly reliable in manual inspection, yet still miss most known vulnerabilities in practice.
Overall, these results capture scanner behavior on confirmed real-world vulnerabilities and provide practical evidence of their recall differences in realistic settings.

\ttbf{False-Positive And False-Negative Causes}
We analyze the root causes of both false positives and false negatives by examining scanner design mechanisms.

\textit{\textbf{Static source-code scanners}} (MCPScan, A.I.G (static)) achieve relatively low false negatives due to their ability to inspect the implementation logic. However, MCPScan reports risks based on potential data-flow semantics without considering runtime exploitability, thereby overestimating risk, while A.I.G (static) relies on LLMs for cross-file semantic analysis and may produce incorrect or even hallucinated data-flow paths in complex codebases. Without runtime constraints, these approaches cannot accurately capture the true attack surface, and their remaining false negatives are primarily due to unstable LLM reasoning.
\textit{\textbf{Metadata-based scanners}} (Agent-Scan, MCP-Scanner, et al.) rely solely on tool descriptions and schemas, lacking visibility into implementation and execution paths. Their detection is typically driven by keyword or semantic matching and further amplified by LLM-based inference, which tends to misinterpret capabilities as vulnerabilities, leading to high false-positive rates; at the same time, the absence of execution semantics prevents them from identifying real vulnerabilities, resulting in significant false negatives in CVE scenarios.
\textit{\textbf{LLM-agent-based dynamic scanners}} (A.I.G (dynamic), MCPSafetyScanner) attempt to validate risks through adversarial probing, but their results are undermined by reasoning instability and flawed validation processes: models often assume strong attacker capabilities and infer high-risk scenarios, while generated probes are limited in number and effectiveness, frequently failing during execution; nevertheless, failed validation may still be interpreted as evidence of risk. In addition, sensitivity to prompts, stochasticity, and timeouts further reduces accuracy, and limited probe coverage leads to missed vulnerabilities that require specific execution paths.
\textit{\textbf{Rule-based scanners}} (nova-proximity, mcp-gateway, mcp-armor) depend on predefined patterns without modeling MCP-specific attack semantics, making their performance highly dependent on rule quality: loose rules trigger superficial pattern matches and false positives, while incomplete rules fail to detect emerging vulnerabilities, causing false negatives.

For example, in \textit{ecommerce-store-mcp}~\cite{ecommerce_store_mcp}, source-aware analysis identifies unsafe external inputs reaching the model, whereas metadata-only scanners miss the issue because the tool description appears benign; conversely, some scanners report credential leakage solely due to the presence of a field named “token” in the schema, even though runtime inspection shows that it is only an input placeholder with no sensitive data access.
Overall, both false positives and false negatives stem from a systematic mismatch between the evidence used by scanners (metadata, static data flows, or LLM reasoning) and the actual runtime semantics of MCP servers, which causes scanners to conflate capability, patterns, or hypothetical reasoning with concrete, exploitable vulnerabilities.

\begin{figure}[t]
\centering
\includegraphics[width=0.9\linewidth]{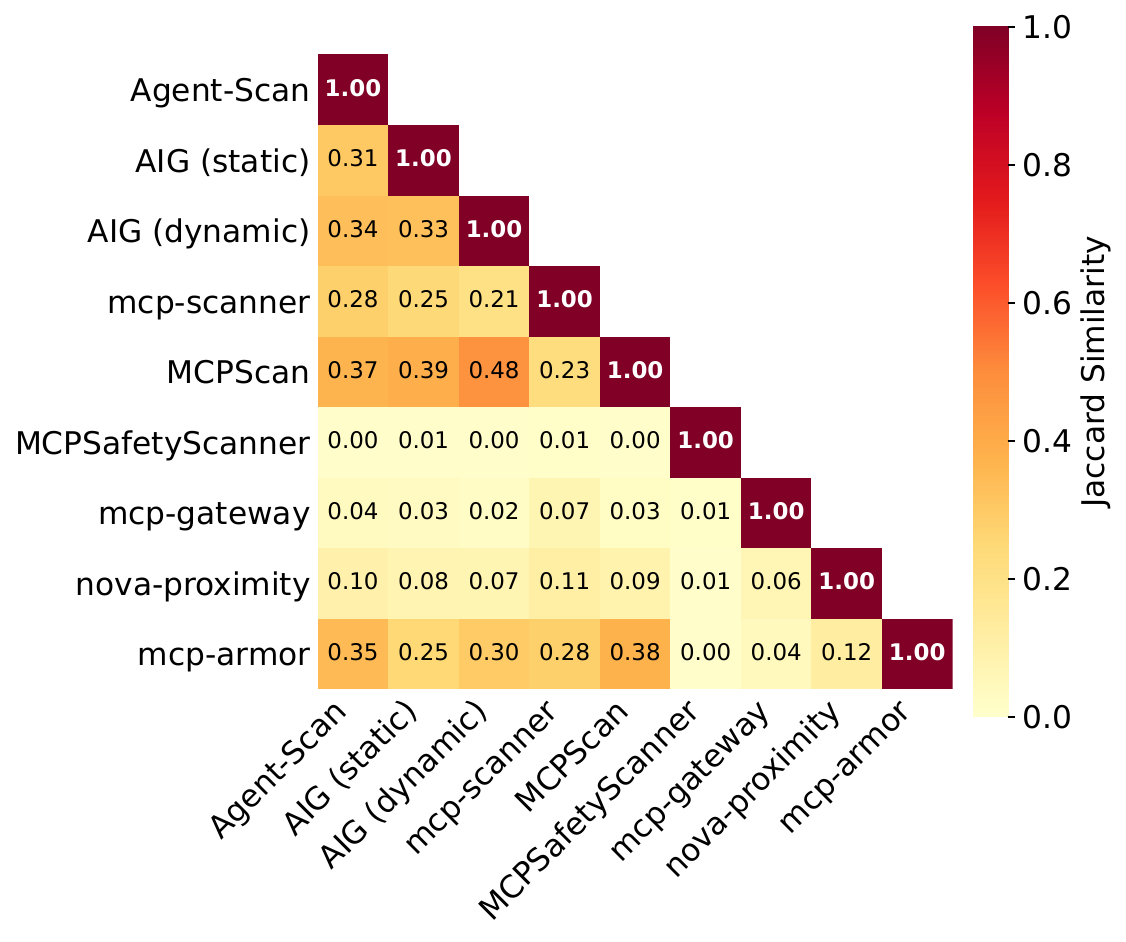}
\caption{Pairwise Jaccard Similarity among All Scanners on the \emph{Any Risk} Criterion, i.e., the Fraction of MCP Servers for Which Both Scanners Report at Least One Vulnerability.}
\Description{Heatmap showing pairwise Jaccard similarity among eight scanners under the Any Risk criterion.}
\label{fig:jaccard_any_risk}
\end{figure}

\ttbf{Cross-Scanner Consistency}
Furthermore, to quantify the agreement across scanners, we measure pairwise Jaccard similarity at two levels.
We choose Jaccard because each scanner output can be represented as a set of flagged servers, and the metric measures relative set overlap rather than absolute alert volume; this follows prior scanner-consistency analysis that uses Jaccard to compare tool-output overlap~\cite{churakova2025vexed}.
Global overlap calculates whether a scanner flags a server for any risk, and category-level overlap evaluates the agreement within specific vulnerability types. For each scanner pair, both measures are computed only over servers successfully processed by both scanners.
Our analysis reveals a severe lack of consensus across the scanner ecosystem.
At the global level, the average pairwise Jaccard similarity is only 15.66\% (Figure~\ref{fig:jaccard_any_risk}).
Even the highest overlap, between A.I.G (dynamic) and MCPScan, reaches just 47.80\%.
This inconsistency becomes even more pronounced at the category level (Appendix~\ref{appendix:consistency}, Figures~\ref{fig:jaccard_prompt}--\ref{fig:jaccard_cred}).
The average similarity drops precipitously across all types: Prompt Injection (5.98\%), Command Execution (5.39\%), and Data Leakage (3.36\%),
indicating that scanners rarely agree on specific vulnerability types.

\finding{
Current MCP scanners exhibit substantial false positives and false negatives, making their outputs unreliable as ground truth for risk assessment.
}

\subsection{Public Query Interface for Scanner Reports}

Our analysis shows that scanner reports are inconsistent and do not reliably reflect actual vulnerabilities.
Meanwhile, MCP servers are frequently copied, renamed, and republished across different marketplaces, making it difficult for users to
assess the security of a given MCP server instance.
As a supporting service, we provide a public query interface\footnote{\url{https://security.fudan.edu.cn/zoo/risk-monitor}} over the normalized scanner reports.

The interface supports two matching modes. First, source-based matching identifies an MCP server by repository URL or source code
when the exact implementation is known.
Second, fuzzy matching supports partial or renamed servers by comparing server name, tool list, descriptions, and input schemas against MCPZoo entries.
For each query, the interface returns matched MCPZoo entities, scanner-reported risks, cross-scanner agreement, and validation status when available.
Importantly, scanner outputs are presented as potential risks rather than confirmed vulnerabilities, in line with our validation results.
This design helps users interpret scanner outputs more cautiously and supports practical risk assessment in the MCP ecosystem.

\section{Discussion}

\subsection{Security Implication}

Weak deployment practices, high-capability tool exposure, and template-driven replication jointly amplify existing attack opportunities in the MCP ecosystem. Market overlap makes the same server implementation visible through multiple discovery channels, so a vulnerable or weakly maintained server can reach users across different MCP markets. Duplication further turns one implementation flaw into many deployable instances: when forks, mirrors, or template-derived replicas preserve the same authentication assumptions, tool descriptions, or unsafe execution paths, the same bug can be inherited by many servers. For example, hundreds of near-identical \texttt{remote-mcp-server-authless} replicas indicate that an authless template can become a repeated deployment pattern rather than an isolated project choice, expanding the number of entry points that attackers or unsafe agents can target.

These issues have important security implications for interpreting scanner reports: inaccurate or inconsistent scanner outputs can mislead both users and developers, either by over-reporting unverified risks or by failing to identify known vulnerabilities, thereby creating a distorted view of the system’s true security posture, where high alert rates may reflect duplicated templates, repeated metadata, or scanner heuristics rather than confirmed vulnerabilities, while low or inconsistent alerts do not guarantee safety. Thus, scanner reports are useful triage signals, but treating them as direct evidence of ecosystem insecurity or security can lead to misplaced effort and missed real vulnerabilities.

Our work provides a scalable foundation for addressing these challenges. MCPZoo enables systematic, runtime-aware evaluation of MCP servers and serves as a shared testbed for developing more reliable security analysis methods. Its public query interface further supports practical risk triage by exposing scanner agreement and validation results, facilitating more informed security assessment in the MCP ecosystem. By connecting overlap, duplication, deployment behavior, tool capabilities, and validated scanner evidence, MCPZoo helps users distinguish structural risk factors from confirmed vulnerabilities and helps scanner developers evaluate whether their tools remain reliable across repeated and variant MCP servers.

\subsection{Limitation}

\ttbf{Deployment Bias}
While MCPZoo significantly expands the set of MCP servers that support dynamic analysis, servers with additional environmental requirements, such as proprietary API credentials or external hardware drivers,
may remain undeployed. This may bias the dataset toward projects that are easier to deploy automatically.

\ttbf{Selection of Available Security Scanners}
Our evaluation focuses on publicly available and widely used MCP scanners, covering diverse analysis strategies under current practice. It may not include emerging methods as the ecosystem evolves. In addition, results depend on scanner versions and model configurations; improvements in scanners or LLMs may further enhance detection performance.

\ttbf{Scale of Ground-Truth Dataset}
Our ground truth combines manual validation and publicly disclosed CVEs, covering representative cases under current MCP practices. As the MCP ecosystem is still evolving, broader benchmark sources (e.g., exploit scenarios or larger confirmed vulnerability sets) remain limited; incorporating them would further strengthen evaluation and is an important direction for future work.

\section{Related Work}
\ttbf{Attack and Defense in the MCP Ecosystem}
As the MCP ecosystem expands, researchers have begun to investigate its security implications.
For \textit{Attacks}, prior work shows that manipulating tool metadata can reliably steer agent tool selection or behavior mode~\cite{wang2025mpma,mo2026attractive,faghih-etal-2025-tool,zhan2026adversarialenvironmentsmisleadagentic}. Recent work further shows that seemingly benign MCP servers can be composed into cross-tool attacks that enable unintended information flow and data exfiltration~\cite{zhao2025mindserversystematicstudy,croce2025trivialtrojansminimalmcp}. Complementary surveys summarize broader attack surfaces in MCP-enabled agent ecosystems~\cite{song2025protocolunveilingattackvectors, FERRAG2026353}.
For \textit{Defenses}, works have begun to mitigate MCP risks through access control, protocol hardening, and runtime monitoring. Prior work constrains agent capabilities through task-aware permission control and safety-oriented protocol refinements~\cite{cai2025grants,jing-etal-2025-mcip}. Other efforts focus on systematic security evaluation and runtime detection of MCP threats~\cite{yang2025mcpsecbenchsystematicsecuritybenchmark,shi2025quantifyingconversationdriftmcp}. Recent work further extends these efforts with automated threat intelligence pipelines for continuous MCP risk tracking and analysis~\cite{shen2026mcpthreathiveautomatedthreatintelligence}.
While these studies establish essential threat models and defensive mechanisms, they remain limited to theoretical analysis or small-scale validation. There is a lack of empirical evidence regarding how these security issues manifest and impact the broader, real-world MCP ecosystem at runtime.
In contrast, our work complements these studies with ecosystem-scale runtime measurement, showing how such risks appear across real MCP servers rather than only in threat models or small validation settings.

\ttbf{The Measurement of MCP Ecosystem}
Recent studies measure the MCP ecosystem from two perspectives: static inspection and runtime evaluation.
\textit{Static studies} inspect MCP artifacts without deployment.
Some work characterizes the ecosystem at scale, including MCP architecture, lifecycle, repository distribution, tool domains, and server-level risks across public registries and open-source repositories~\cite{hou2025modelcontextprotocolmcp, li2025understandingsecurityissuesmodel, stein2026aiagentsusedevidence}. Another works focus on security-oriented static reasoning, including threat taxonomies, MCP-specific scanners, CWE/CAPEC-based risk assessment, and source-code vulnerability detection~\cite{guo2025systematicanalysismcpsecurity, hasan2026modelcontextprotocolmcp, kumar2026mcpinsosriskassessmentframework}.
While these approaches scale well, they cannot verify deployment feasibility or runtime behavior.
Our study addresses this gap by separating collected ecosystem scale from the subset that can actually be deployed, invoked, and evaluated at runtime.
\textit{Dynamic studies} deploy MCP servers or agent-tool environments and observe concrete interactions, making them better suited for evaluating runtime behavior. Existing work has examined emergent agent misuse, tool poisoning, prompt-injection-style attacks, and executable MCP security benchmarks in curated settings~\cite{noever2025servantstalkerpredatorhonest, wang2025mcptoxbenchmarktoolpoisoning, debenedetti2024agentdojodynamicenvironmentevaluate, yang2025mcpsecbenchsystematicsecuritybenchmark, zhang2026mcpsecuritybenchmsb,tiwari2025modelcontextprotocolvision}.
However, they are typically limited in scale and coverage.
MCPZoo bridges this gap by enabling automated large-scale deployment and interactive measurement of real-world MCP servers.
Rather than focusing on curated attacks or selected server sets, our measurement starts from public ecosystem collections and reports deployability, tool exposure, scanner outputs, and validation outcomes at larger scale.
Our earlier MCPZoo preprint~\cite{wu2025mcpzoo} introduced the initial dataset and automated deployment framework. This article extends that work with an expanded runtime corpus, ecosystem characterization, and a large-scale reliability evaluation and validation of MCP security scanners.

\section{Conclusion}

We present MCPZoo, the first large-scale dataset that enables dynamic, runtime security evaluation of real-world MCP servers.
Our study reveals that the MCP ecosystem not only exhibits structural risks, but also that scanner-reported risks do not reliably indicate true vulnerabilities.
These findings highlight a critical gap between large-scale security measurement and trustworthy risk assessment.
MCPZoo bridges this gap by enabling systematic validation of scanner outputs and supporting practical risk triage.

\bibliographystyle{ACM-Reference-Format}
\bibliography{reference}

@misc{mcpworld,
  title        = {{MCP World}},
  author       = {{MCP World}},
  key          = {MCP World},
  howpublished = {\url{https://www.mcpworld.com/}},
  year         = {2025},
  note         = {Accessed: 2025-12}
}

@misc{mcpso,
  title        = {{Mcp.so}},
  author       = {{Mcp.so}},
  key          = {Mcp.so},
  howpublished = {\url{https://mcp.so}},
  year         = {2025},
  note         = {Accessed: 2025-12}
}

@misc{mcprepository,
  title        = {{MCP Repository}},
  author       = {{MCP Repository}},
  key          = {MCP Repository},
  howpublished = {\url{https://mcprepository.com/}},
  year         = {2025},
  note         = {Accessed: 2025-12}
}

@misc{aibasemcp,
  title        = {{AIbase MCP}},
  author       = {{AIbase}},
  key          = {AIbase MCP},
  howpublished = {\url{https://mcp.aibase.com}},
  year         = {2025},
  note         = {Accessed: 2025-12}
}

@misc{mcpstore,
  title        = {{MCP Store}},
  author       = {{MCP Store}},
  key          = {MCP Store},
  howpublished = {\url{https://mcpstore.co}},
  year         = {2025},
  note         = {Accessed: 2025-12}
}

@misc{pulsemcp,
  title        = {{Pulse MCP}},
  author       = {{Pulse MCP}},
  key          = {Pulse MCP},
  howpublished = {\url{https://www.pulsemcp.com/servers}},
  year         = {2025},
  note         = {Accessed: 2025-12}
}

@misc{mcpmarket,
  title        = {{MCP Market}},
  author       = {{MCP Market}},
  key          = {MCP Market},
  howpublished = {\url{https://mcpmarket.com/}},
  year         = {2025},
  note         = {Accessed: 2025-12}
}

@misc{smithery,
  title        = {{Smithery}},
  author       = {{Smithery}},
  key          = {Smithery},
  howpublished = {\url{https://smithery.ai}},
  year         = {2025},
  note         = {Accessed: 2025-12}
}

@misc{npm_registry,
  title        = {npm Registry},
  author       = {{npm, Inc.}},
  howpublished = {\url{https://www.npmjs.com/}},
  year         = {2026},
  note         = {Accessed: 2026-02-03}
}

@misc{pypi_registry,
  title        = {Python Package Index (PyPI)},
  author       = {{Python Software Foundation}},
  howpublished = {\url{https://pypi.org/}},
  year         = {2026},
  note         = {Accessed: 2026-02-03}
}

@misc{Tencent_AI-Infra-Guard_2025,
  author={{Tencent Zhuque Lab}},
  title={{AI-Infra-Guard: A Comprehensive, Intelligent, and Easy-to-Use AI Red Teaming Platform}},
  year={2025},
  howpublished={GitHub repository},
  url={https://github.com/Tencent/AI-Infra-Guard}
}

@misc{mcp_scan2025,
  title        = {Agent-Scan: A Static Analysis Tool for Detecting Security Issues in MCP Servers},
  author       = {{Snyk}},
  year         = {2025},
  howpublished = {\url{https://github.com/snyk/agent-scan}},
  note         = {Accessed: 2025-12}
}

@misc{guo2025mcpecosystem,
  title         = {A Measurement Study of Model Context Protocol Ecosystem},
  author        = {Guo, Hechuan and Hao, Yongle and Zhang, Yue and Xu, Minghui and Lv, Peizhuo and Chen, Jiezhi and Cheng, Xiuzhen},
  year          = {2025},
  eprint        = {2509.25292},
  archivePrefix = {arXiv},
  primaryClass  = {cs.CY}
}

@article{liu2024lost,
  title     = {Lost in the Middle: How Language Models Use Long Contexts},
  author    = {Liu, Nelson F. and Lin, Kevin and Hewitt, John and Paranjape, Ashwin and Bevilacqua, Michele and Petroni, Fabio and Liang, Percy},
  journal   = {Transactions of the Association for Computational Linguistics},
  volume    = {12},
  year      = {2024},
  address   = {Cambridge, MA},
  publisher = {MIT Press},
  url       = {https://aclanthology.org/2024.tacl-1.9/},
  doi       = {10.1162/tacl_a_00638},
  pages     = {157--173}
}

@inproceedings{gao2024insights,
  title     = {Insights into {LLM} Long-Context Failures: When Transformers Know but Don{'}t Tell},
  author    = {Gao, Muhan  and Lu, TaiMing  and Yu, Kuai  and Byerly, Adam  and Khashabi, Daniel},
  editor    = {Al-Onaizan, Yaser  and Bansal, Mohit  and Chen, Yun-Nung},
  booktitle = {Findings of the Association for Computational Linguistics: EMNLP 2024},
  month     = nov,
  year      = {2024},
  address   = {Miami, Florida, USA},
  publisher = {Association for Computational Linguistics},
  url       = {https://aclanthology.org/2024.findings-emnlp.447/},
  doi       = {10.18653/v1/2024.findings-emnlp.447},
  pages     = {7611--7625}
}

@misc{mcp-scanner-cisco,
  title        = {{mcp-scanner}: MCP Security Scanner},
  author       = {{Cisco AI Defense}},
  year         = {2025},
  howpublished = {\url{https://github.com/cisco-ai-defense/mcp-scanner}},
  note         = {GitHub repository},
}

@misc{mcpscan-antgroup,
  title        = {{MCPScan}: Security Analysis Tool for MCP Servers},
  author       = {{Ant Group}},
  year         = {2025},
  howpublished = {\url{https://github.com/antgroup/MCPScan}},
  note         = {GitHub repository},
}

@misc{zhao2025mindserversystematicstudy,
      title={Parasites in the Toolchain: A Large-Scale Analysis of Attacks on the MCP Ecosystem}, 
      author={Shuli Zhao and Qinsheng Hou and Zihan Zhan and Yanhao Wang and Yuchong Xie and Yu Guo and Libo Chen and Shenghong Li and Zhi Xue},
      year={2026},
      eprint={2509.06572},
      archivePrefix={arXiv},
      primaryClass={cs.CR},
      url={https://arxiv.org/abs/2509.06572}, 
}

@misc{anthropic2024mcp,
  title = {{Model Context Protocol (MCP) Specification}},
  author = {Anthropic},
  year = {2024},
  url = {https://modelcontextprotocol.io/docs/learn/architecture},
  note = {Accessed: 2025-12}
}

@misc{guo2025systematicanalysismcpsecurity,
      title={Systematic Analysis of MCP Security}, 
      author={Yongjian Guo and Puzhuo Liu and Wanlun Ma and Zehang Deng and Xiaogang Zhu and Peng Di and Xi Xiao and Sheng Wen},
      year={2025},
      eprint={2508.12538},
      archivePrefix={arXiv},
      primaryClass={cs.CR},
      url={https://arxiv.org/abs/2508.12538}, 
}

@misc{mcp-registry,
  title        = {Model Context Protocol Registry},
  author       = {{Model Context Protocol Community}},
  year         = {2025},
  howpublished = {\url{https://registry.modelcontextprotocol.io/}},
  note         = {Official MCP server registry},
}

@misc{anthropic-mcp-donation,
  title        = {Donating the Model Context Protocol and Establishing of the Agentic {AI} Foundation},
  author       = {{Anthropic}},
  year         = {2025},
  howpublished = {\url{https://www.anthropic.com/news/donating-the-model-context-protocol-and-establishing-of-the-agentic-ai-foundation}},
  note         = {Official announcement},
}

@misc{fastapi,
  title        = {FastAPI},
  author       = {FastAPI},
  year         = {2025},
  howpublished = {\url{https://fastapi.tiangolo.com/}},
  note         = {Official website},
}

@misc{supergateway,
  title        = {Supergateway},
  author       = {{supercorp-ai}},
  year         = {2025},
  howpublished = {\url{https://github.com/supercorp-ai/supergateway}},
  note         = {GitHub repository},
}

@article{wang2025mpma,
  title={Mpma: Preference manipulation attack against model context protocol},
  author={Wang, Zihan and Zhang, Rui and Liu, Yu and Fan, Wenshu and Jiang, Wenbo and Zhao, Qingchuan and Li, Hongwei and Xu, Guowen},
  journal={Proceedings of the AAAI Conference on Artificial Intelligence},
  volume={40},
  number={42},
  pages={35838--35846},
  year={2026}
}

@misc{song2025protocolunveilingattackvectors,
      title={Beyond the Protocol: Unveiling Attack Vectors in the Model Context Protocol (MCP) Ecosystem}, 
      author={Hao Song and Yiming Shen and Wenxuan Luo and Leixin Guo and Ting Chen and Jiashui Wang and Beibei Li and Xiaosong Zhang and Jiachi Chen},
      year={2025},
      eprint={2506.02040},
      archivePrefix={arXiv},
      primaryClass={cs.CR},
      url={https://arxiv.org/abs/2506.02040}, 
}

@misc{croce2025trivialtrojansminimalmcp,
      title={Trivial Trojans: How Minimal MCP Servers Enable Cross-Tool Exfiltration of Sensitive Data}, 
      author={Nicola Croce and Tobin South},
      year={2025},
      eprint={2507.19880},
      archivePrefix={arXiv},
      primaryClass={cs.CR},
      url={https://arxiv.org/abs/2507.19880}, 
}

@misc{yang2025mcpsecbenchsystematicsecuritybenchmark,
      title={MCPSecBench: A Systematic Security Benchmark and Playground for Testing Model Context Protocols}, 
      author={Yixuan Yang and Daoyuan Wu and Yufan Chen},
      year={2025},
      eprint={2508.13220},
      archivePrefix={arXiv},
      primaryClass={cs.CR},
      url={https://arxiv.org/abs/2508.13220}, 
}

@misc{shi2025quantifyingconversationdriftmcp,
      title={Quantifying Conversation Drift in MCP via Latent Polytope}, 
      author={Haoran Shi and Hongwei Yao and Shuo Shao and Shaopeng Jiao and Ziqi Peng and Zhan Qin and Cong Wang},
      year={2025},
      eprint={2508.06418},
      archivePrefix={arXiv},
      primaryClass={cs.CL},
      url={https://arxiv.org/abs/2508.06418}, 
}

@misc{hou2025modelcontextprotocolmcp,
      title={Model Context Protocol (MCP): Landscape, Security Threats, and Future Research Directions}, 
      author={Xinyi Hou and Yanjie Zhao and Shenao Wang and Haoyu Wang},
      year={2025},
      eprint={2503.23278},
      archivePrefix={arXiv},
      primaryClass={cs.CR},
      url={https://arxiv.org/abs/2503.23278}, 
}

@misc{li2025understandingsecurityissuesmodel,
      title={A First Look at the Security Issues in the Model Context Protocol Ecosystem}, 
      author={Xiaofan Li and Xing Gao},
      year={2026},
      eprint={2510.16558},
      archivePrefix={arXiv},
      primaryClass={cs.CR},
      url={https://arxiv.org/abs/2510.16558}, 
}

@misc{noever2025servantstalkerpredatorhonest,
      title={Servant, Stalker, Predator: How An Honest, Helpful, And Harmless (3H) Agent Unlocks Adversarial Skills}, 
      author={David Noever},
      year={2025},
      eprint={2508.19500},
      archivePrefix={arXiv},
      primaryClass={cs.CR},
      url={https://arxiv.org/abs/2508.19500}, 
}

@misc{wang2025mcptoxbenchmarktoolpoisoning,
      title={MCPTox: A Benchmark for Tool Poisoning Attack on Real-World MCP Servers}, 
      author={Zhiqiang Wang and Yichao Gao and Yanting Wang and Suyuan Liu and Haifeng Sun and Haoran Cheng and Guanquan Shi and Haohua Du and Xiangyang Li},
      year={2025},
      eprint={2508.14925},
      archivePrefix={arXiv},
      primaryClass={cs.CR},
      url={https://arxiv.org/abs/2508.14925}, 
}

@misc{ecommerce_store_mcp,
  author = {khesayed},
  title = {ecommerce-store-mcp: A Model Context Protocol Server for E-commerce},
  year = {2025},
  publisher = {GitHub},
  journal = {GitHub repository},
  howpublished = {\url{https://github.com/khesayed/ecommerce-store-mcp}}
}

@misc{mcp_gsuite,
  author = {w-10-m},
  title = {gsuite: An MCP Server for Google Workspace Integration},
  year = {2025},
  publisher = {GitHub},
  journal = {GitHub repository},
  howpublished = {\url{https://github.com/w-10-m/gsuite}}
}

@misc{mcp_nomad,
  author = {kocierik},
  title = {mcp-nomad: A Model Context Protocol Server for HashiCorp Nomad},
  year = {2025},
  publisher = {GitHub},
  journal = {GitHub repository},
  howpublished = {\url{https://github.com/kocierik/mcp-nomad}}
}

@misc{github_rest_api,
  author = {{GitHub}},
  title = {GitHub REST API Documentation},
  year = {2026},
  publisher = {GitHub},
  howpublished = {\url{https://docs.github.com/en/rest}},
  note = {Accessed: 2026-02-06}
}

@misc{mcpsafetyscanner,
  author = {Radosevich, Brandon and Halloran, John},
  title = {{MCP Safety Audit: LLMs with the Model Context Protocol Allow Major Security Exploits}},
  year = {2025},
  eprint = {2504.03767},
  archivePrefix = {arXiv},
  primaryClass = {cs.CR},
  howpublished = {\url{https://www.arxiv.org/abs/2504.03767}},
  note = {MCPSafetyScanner}
}

@misc{mcp-gateway,
  author = {{Lasso Security}},
  title = {{MCP Gateway}},
  year = {2025},
  publisher = {GitHub},
  journal = {GitHub repository},
  howpublished = {\url{https://github.com/lasso-security/mcp-gateway}}
}

@misc{nova-proximity,
  author = {{Nova Hunting}},
  title = {{nova-proximity: MCP and Claude Skill Security Scanner}},
  year = {2026},
  publisher = {GitHub},
  journal = {GitHub repository},
  howpublished = {\url{https://github.com/Nova-Hunting/nova-proximity}}
}

@misc{mcp-armor,
  author = {{Aira Security}},
  title = {{mcp-armor: MCP Configuration Scanner with Client-Aware Security Analysis}},
  year = {2026},
  publisher = {GitHub},
  journal = {GitHub repository},
  howpublished = {\url{https://github.com/aira-security/mcp-armor}}
}

@misc{nvd-cve-2025-53818,
  author = {{National Institute of Standards and Technology}},
  title = {{CVE-2025-53818 Detail}},
  year = {2025},
  howpublished = {\url{https://nvd.nist.gov/vuln/detail/CVE-2025-53818}},
  note = {National Vulnerability Database}
}

@misc{nvd-cve-2025-66580,
  author = {{National Institute of Standards and Technology}},
  title = {{CVE-2025-66580 Detail}},
  year = {2025},
  howpublished = {\url{https://nvd.nist.gov/vuln/detail/CVE-2025-66580}},
  note = {National Vulnerability Database}
}

@misc{nvd-cve-2025-68669,
  author = {{National Institute of Standards and Technology}},
  title = {{CVE-2025-68669 Detail}},
  year = {2025},
  howpublished = {\url{https://nvd.nist.gov/vuln/detail/CVE-2025-68669}},
  note = {National Vulnerability Database}
}

@misc{nvd-cve-2026-22793,
  author = {{National Institute of Standards and Technology}},
  title = {{CVE-2026-22793 Detail}},
  year = {2026},
  howpublished = {\url{https://nvd.nist.gov/vuln/detail/CVE-2026-22793}},
  note = {National Vulnerability Database}
}

@misc{nvd-cve-2026-25546,
  author = {{National Institute of Standards and Technology}},
  title = {{CVE-2026-25546 Detail}},
  year = {2026},
  howpublished = {\url{https://nvd.nist.gov/vuln/detail/CVE-2026-25546}},
  note = {National Vulnerability Database}
}

@misc{nvd-cve-2026-25650,
  author = {{National Institute of Standards and Technology}},
  title = {{CVE-2026-25650 Detail}},
  year = {2026},
  howpublished = {\url{https://nvd.nist.gov/vuln/detail/CVE-2026-25650}},
  note = {National Vulnerability Database}
}

@misc{nvd-cve-2026-27825,
  author = {{National Institute of Standards and Technology}},
  title = {{CVE-2026-27825 Detail}},
  year = {2026},
  howpublished = {\url{https://nvd.nist.gov/vuln/detail/CVE-2026-27825}},
  note = {National Vulnerability Database}
}

@misc{nvd-cve-2026-33946,
  author = {{National Institute of Standards and Technology}},
  title = {{CVE-2026-33946 Detail}},
  year = {2026},
  howpublished = {\url{https://nvd.nist.gov/vuln/detail/CVE-2026-33946}},
  note = {National Vulnerability Database}
}

@misc{nvd-cve-2026-33980,
  author = {{National Institute of Standards and Technology}},
  title = {{CVE-2026-33980 Detail}},
  year = {2026},
  howpublished = {\url{https://nvd.nist.gov/vuln/detail/CVE-2026-33980}},
  note = {National Vulnerability Database}
}

@misc{nvd-cve-2026-39884,
  author = {{National Institute of Standards and Technology}},
  title = {{CVE-2026-39884 Detail}},
  year = {2026},
  howpublished = {\url{https://nvd.nist.gov/vuln/detail/CVE-2026-39884}},
  note = {National Vulnerability Database}
}

@misc{ray2025survey,
  title={A survey on model context protocol: Architecture, state-of-the-art, challenges and future directions},
  author={Ray, Partha Pratim},
  year={2025},
  howpublished={Authorea Preprints},
  publisher={Authorea}
}

@misc{mo2026attractive,
  title={Attractive Metadata Attack: Inducing {LLM} Agents to Invoke Malicious Tools},
  author={Kanghua Mo and Li Hu and Yucheng Long and Zhihao li},
  year={2026},
  howpublished={The Thirty-ninth Annual Conference on Neural Information Processing Systems},
  url={https://openreview.net/forum?id=oLGtPYdRzU}
}

@inproceedings{faghih-etal-2025-tool,
    title = "Tool Preferences in Agentic {LLM}s are Unreliable",
    author = "Faghih, Kazem  and
      Wang, Wenxiao  and
      Cheng, Yize  and
      Bharti, Siddhant  and
      Sriramanan, Gaurang  and
      Balasubramanian, Sriram  and
      Hosseini, Parsa  and
      Feizi, Soheil",
    editor = "Christodoulopoulos, Christos  and
      Chakraborty, Tanmoy  and
      Rose, Carolyn  and
      Peng, Violet",
    booktitle = "Proceedings of the 2025 Conference on Empirical Methods in Natural Language Processing",
    month = nov,
    year = "2025",
    address = "Suzhou, China",
    publisher = "Association for Computational Linguistics",
    url = "https://aclanthology.org/2025.emnlp-main.1060/",
    doi = "10.18653/v1/2025.emnlp-main.1060",
    pages = "20954--20969",
    ISBN = "979-8-89176-332-6"
}

@article{FERRAG2026353,
title = {From prompt injections to protocol exploits: Threats in LLM-powered AI agents workflows},
journal = {ICT Express},
volume = {12},
number = {2},
pages = {353-383},
year = {2026},
issn = {2405-9595},
doi = {https://doi.org/10.1016/j.icte.2025.12.001},
url = {https://www.sciencedirect.com/science/article/pii/S2405959525001997},
author = {Mohamed Amine Ferrag and Norbert Tihanyi and Djallel Hamouda and Leandros Maglaras and Abderrahmane Lakas and Merouane Debbah},
}

@misc{cai2025grants,
  title={Who Grants the Agent Power? Defending Against Instruction Injection via Task-Centric Access Control},
  author={Cai, Yifeng and Wang, Ziming and Deng, Zhaomeng and Yao, Mengyu and Liu, Junlin and Hu, Yutao and Zhang, Ziqi and Guo, Yao and Li, Ding},
  year={2025},
  eprint={2510.26212},
  archivePrefix={arXiv},
  primaryClass={cs.CR}
}

@inproceedings{jing-etal-2025-mcip,
    title = "{MCIP}: Protecting {MCP} Safety via Model Contextual Integrity Protocol",
    author = "Jing, Huihao  and
      Li, Haoran  and
      Hu, Wenbin  and
      Hu, Qi  and
      Heli, Xu  and
      Chu, Tianshu  and
      Hu, Peizhao  and
      Song, Yangqiu",
    editor = "Christodoulopoulos, Christos  and
      Chakraborty, Tanmoy  and
      Rose, Carolyn  and
      Peng, Violet",
    booktitle = "Proceedings of the 2025 Conference on Empirical Methods in Natural Language Processing",
    month = nov,
    year = "2025",
    address = "Suzhou, China",
    publisher = "Association for Computational Linguistics",
    url = "https://aclanthology.org/2025.emnlp-main.62/",
    doi = "10.18653/v1/2025.emnlp-main.62",
    pages = "1177--1194",
    ISBN = "979-8-89176-332-6",
}

@misc{shen2026mcpthreathiveautomatedthreatintelligence,
      title={MCPThreatHive: Automated Threat Intelligence for Model Context Protocol Ecosystems}, 
      author={Yi Ting Shen and Kentaroh Toyoda and Alex Leung},
      year={2026},
      eprint={2604.13849},
      archivePrefix={arXiv},
      primaryClass={cs.CR},
      url={https://arxiv.org/abs/2604.13849}, 
}

@misc{zhan2026adversarialenvironmentsmisleadagentic,
      title={How Adversarial Environments Mislead Agentic AI?}, 
      author={Zhonghao Zhan and Huichi Zhou and Zhenhao Li and Peiyuan Jing and Krinos Li and Hamed Haddadi},
      year={2026},
      eprint={2604.18874},
      archivePrefix={arXiv},
      primaryClass={cs.AI},
      url={https://arxiv.org/abs/2604.18874}, 
}

@misc{hasan2026modelcontextprotocolmcp,
      title={Model Context Protocol (MCP) at First Glance: Studying the Security and Maintainability of MCP Servers}, 
      author={Mohammed Mehedi Hasan and Hao Li and Emad Fallahzadeh and Gopi Krishnan Rajbahadur and Bram Adams and Ahmed E. Hassan},
      year={2026},
      eprint={2506.13538},
      archivePrefix={arXiv},
      primaryClass={cs.SE},
      url={https://arxiv.org/abs/2506.13538}, 
}

@misc{stein2026aiagentsusedevidence,
      title={How are AI agents used? Evidence from 177,000 MCP tools}, 
      author={Merlin Stein},
      year={2026},
      eprint={2603.23802},
      archivePrefix={arXiv},
      primaryClass={cs.CY},
      url={https://arxiv.org/abs/2603.23802}, 
}

@misc{kumar2026mcpinsosriskassessmentframework,
      title={MCP-in-SoS: Risk assessment framework for open-source MCP servers}, 
      author={Pratyay Kumar and Miguel Antonio Guirao Aguilera and Srikathyayani Srikanteswara and Satyajayant Misra and Abu Saleh Md Tayeen},
      year={2026},
      eprint={2603.10194},
      archivePrefix={arXiv},
      primaryClass={cs.CR},
      url={https://arxiv.org/abs/2603.10194}, 
}

@misc{debenedetti2024agentdojodynamicenvironmentevaluate,
      title={AgentDojo: A Dynamic Environment to Evaluate Prompt Injection Attacks and Defenses for LLM Agents}, 
      author={Edoardo Debenedetti and Jie Zhang and Mislav Balunović and Luca Beurer-Kellner and Marc Fischer and Florian Tramèr},
      year={2024},
      eprint={2406.13352},
      archivePrefix={arXiv},
      primaryClass={cs.CR},
      url={https://arxiv.org/abs/2406.13352}, 
}

@misc{zhang2026mcpsecuritybenchmsb,
      title={MCP Security Bench (MSB): Benchmarking Attacks Against Model Context Protocol in LLM Agents}, 
      author={Dongsen Zhang and Zekun Li and Xu Luo and Xuannan Liu and Peipei Li and Wenjun Xu},
      year={2026},
      eprint={2510.15994},
      archivePrefix={arXiv},
      primaryClass={cs.CR},
      url={https://arxiv.org/abs/2510.15994}, 
}

@misc{tiwari2025modelcontextprotocolvision,
      title={Model Context Protocol for Vision Systems: Audit, Security, and Protocol Extensions}, 
      author={Aditi Tiwari and Akshit Bhalla and Darshan Prasad},
      year={2025},
      eprint={2509.22814},
      archivePrefix={arXiv},
      primaryClass={cs.CR},
      url={https://arxiv.org/abs/2509.22814}, 
}

@misc{churakova2025vexed,
      title={VEXed: Does {VEX} Itself Need Security Fixes?},
      author={Churakova, Olga and Ekstedt, Mathias and Lenarduzzi, Valentina},
      year={2025},
      eprint={2503.14388},
      archivePrefix={arXiv},
      primaryClass={cs.CR},
      url={https://arxiv.org/abs/2503.14388},
}

@misc{wu2025mcpzoo,
  title={MCPZoo: A Large-Scale Dataset of Runnable Model Context Protocol Servers for AI Agent},
  author={Wu, Mengying and Chen, Pei and Hong, Geng and An, Baichao and Chen, Jinsong and Wan, Binwang and Pan, Xudong and Dai, Jiarun and Yang, Min},
  year={2025},
  eprint={2512.15144},
  archivePrefix={arXiv},
  primaryClass={cs.CR},
  url={https://arxiv.org/abs/2512.15144}
}

\appendix
\begin{table*}[t]
\centering
\caption{Execution Setup for the Evaluated MCP Security Scanners.}
\label{tab:scanner_setup}
\resizebox{\textwidth}{!}{
\begin{tabular}{l l l l l}
\toprule
\textbf{Scanner} & \textbf{Mode} & \textbf{Input} & \textbf{Intermediate Artifacts} & \textbf{Analysis Engine} \\

\midrule
Agent-Scan~\cite{mcp_scan2025}
& Dynamic
& Live MCP server endpoint
& Tool descriptions
& Local rules + Remote Snyk API \\
A.I.G (static)~\cite{Tencent_AI-Infra-Guard_2025}
& Static
& Source code path
& Source code
& LLM \\
A.I.G (dynamic)~\cite{Tencent_AI-Infra-Guard_2025}
& Dynamic
& Live MCP server endpoint
& Tool invocation logs
& LLM \\
MCP-Scanner~\cite{mcp-scanner-cisco}
& Dynamic
& Live MCP server endpoint
& Tool descriptions
& YARA rules + LLM \\
MCPScan~\cite{mcpscan-antgroup}
& Static
& Source code path
& Source code, data flows, tool descriptions
& Semgrep + LLM \\
MCPSafetyScanner~\cite{mcpsafetyscanner}
& Dynamic
& Live MCP server endpoint
& External knowledge, tool invocation logs
& LLM + external search tools \\
mcp-gateway~\cite{mcp-gateway}
& Dynamic
& Live MCP server endpoint
& External metadata, tool descriptions
& Search engine + local rules \\
nova-proximity~\cite{nova-proximity}
& Dynamic
& Live MCP server endpoint
& Tool descriptions, prompts, resources
& NOVA rules + LLM \\
mcp-armor~\cite{mcp-armor}
& Dynamic
& Live MCP server endpoint
& Tool descriptions, prompts, resources
& Local rules + FT-Llama-Prompt-Guard-2 \\
\bottomrule
\end{tabular}
}
\end{table*}

\section{Ethical Considerations}

We structure our ethical considerations by linking a stakeholder-based analysis to the impacts generated during two distinct phases of this work: the \textit{research process} (data collection, construction, and analysis) and the \textit{publication and future use} of MCPZoo. We first identify the relevant stakeholders, then discuss the potential benefits and harms of our study, followed by the mitigation measures we adopt. We conclude with a justification for conducting and publishing this research.

\ttbf{Stakeholders}
This work involves the following primary stakeholder groups:
(1) \textit{MCP Server Developers}:
Developers whose MCP server code is collected, analyzed, built, and executed as part of MCPZoo.
(2) \textit{MCP Server Markets and Registries}:
Platforms that host MCP servers and associated metadata, which are crawled and integrated into our dataset.
(3) \textit{Security Scanner Developers}:
Developers of third-party security scanners whose scanners are used and evaluated in our study.
(4) \textit{The Security Research Community}:
Researchers who may use MCPZoo as a resource for studying the MCP ecosystem.
(5) \textit{The Research Team}:
Our own research team, which executes untrusted code and bears operational and security risks during experimentation.

\ttbf{Impacts of the Research}
The construction and publication of MCPZoo have both positive and negative impacts on the stakeholders.

\noindent\textit{Potential Benefits.}
MCPZoo provides the MCP ecosystem with a large-scale dataset that supports dynamic analysis of MCP servers. By enabling runtime execution, protocol-level interaction, and systematic evaluation, MCPZoo facilitates research on dynamic security scanning, vulnerability validation, defense testing, and ecosystem-level measurement. This capability helps reduce biases introduced by prior studies that rely solely on static analysis or small, manually curated samples, and supports more reproducible and evidence-based security research.

\noindent\textit{Potential Harms.}
Despite these benefits, our work also entails several potential risks:
(1) Collecting and analyzing third-party MCP server code may involve sensitive configurations or private information embedded in repositories.
(2) Crawling market platforms may impose additional load.
(3) Executing MCP servers from untrusted sources introduces risks to local infrastructure, including resource abuse or exploitation.
(4) Some security scanners, such as Agent-Scan, require interaction with remote services during analysis, which may generate additional load on their servers.
(5) By lowering the barrier to building and running realistic MCP services, MCPZoo may also lower the barrier for adversaries to experiment with or test attacks.

\ttbf{Mitigations}
To address the risks outlined above, we adopt multiple mitigation measures during both the experimental phase and future deployment.

\noindent\textit{Mitigations During the Research Process.}
(1) We collect MCP server code and metadata exclusively from publicly accessible sources, including open repositories and market-provided metadata. Our data collection adheres to robots.txt rules, and applies conservative crawling rates to avoid excessive load.
(2) We do not collect or publish authentication tokens, API keys, or other sensitive configuration data from the repositories. During testing, any required credentials are replaced with locally deployed test keys or empty placeholders.
(3) All MCP servers are executed within containerized sandbox environments using the principle of least privilege. We restrict CPU and memory usage for each container and retain execution logs to support exception detection and auditing.
(4) Security scanners are used strictly according to their documented and intended usage. All scanners are deployed locally and connected to locally hosted large language models. For the only scanner requiring remote interaction (Agent-Scan~\cite{mcp_scan2025}), we strictly control request rates and do not perform any form of stress testing or adversarial interaction with the provider's infrastructure.

\noindent\textit{Mitigations for Future Use and Deployment.}
In future releases, MCPZoo will provide access interfaces under controlled conditions. Access will be restricted to legitimate, non-commercial academic use, requiring applicants to submit usage justifications and identity information. To protect MCPZoo's infrastructure from abuse, we plan to enforce authentication, isolation, and rate-limiting mechanisms (e.g., request frequency limits) on exposed interfaces.

\ttbf{Decision and Justification}
In summary, this work addresses a critical challenge in the MCP ecosystem: the lack of large-scale support for dynamic security analysis. Through careful experimental design and responsible mitigation measures, we minimize potential harms to MCP server developers, MCP server markets, scanner providers, and the research team. Looking forward, we will continue to support academic research on MCP security under controlled access, ensuring that MCPZoo serves the research community while safeguarding infrastructure, stakeholders, and public interest.
Above analyses therefore converge on the same conclusion: proceeding with and publishing this research is ethically justified. We further believe that withholding this work would perpetuate existing blind spots in MCP security, which poses a longer-term risk to the ecosystem.

\section{Scanner Analysis Logic}
\label{sec:scanner_logic}

\begin{table}[t]
\centering
\caption{Mapping of Scanner-Specific Risk Labels to Unified Categories.}
\label{tab:category_mapping}
\resizebox{\columnwidth}{!}{
\begin{tabular}{l l p{5.5cm}}
\toprule
\textbf{Unified Category} & \textbf{Scanner} & \textbf{Original Labels} \\
\midrule
\multirow{9}{*}{\makecell[l]{\textbf{Prompt }\\ \textbf{Injection}}}
& Agent-Scan       & E001, E003 \\
& MCP-Scanner      & \textsc{Prompt Injection}, \textsc{Indirect Prompt Injection}, \textsc{Tool Poisoning}, \textsc{Tool Shadowing}, Tool Metadata Pollution \\
& A.I.G            & MCP03, MCP06, Tool Poisoning, Tool Shadowing Attack, Name Confusion \\
& MCPScan          & Indirect Prompt Injection, Tool Metadata Pollution \\
& MCPSafetyScanner & Prompt Injection \\
& mcp-gateway      & hidden\_instructions \\
& nova-proximity   & Prompt Injection \\
& mcp-armor        & Prompt Injection, Tool Name Ambiguity \\
\midrule
\multirow{7}{*}{\makecell[l]{\textbf{Command}\\ \textbf{Execution}}}
& Agent-Scan       & TF002 \\
& MCP-Scanner      & \textsc{Code Execution}, \textsc{Injection Attack}, \textsc{System Manipulation} \\
& A.I.G            & MCP05 \\
& MCPScan          & Tool Poisoning \& Malicious Code Snippets \\
& MCPSafetyScanner & Command Execution, Destructive Action \\
& mcp-gateway      & sensitive\_actions \\
& mcp-armor        & Command Injection \\
\midrule
\multirow{6}{*}{\makecell[l]{\textbf{Data}\\ \textbf{Leakage}}}
& Agent-Scan       & TF001 \\
& MCP-Scanner      & \textsc{Data Exfiltration}, \textsc{Credential Harvesting} \\
& A.I.G            & MCP01 \\
& MCPScan          & detect-hardcoded-secrets-py/js \\
& MCPSafetyScanner & Secret Exposure, Sensitive Data Exposure \\
& mcp-gateway      & sensitive\_files \\
\midrule
\end{tabular}
}
\begin{minipage}{\columnwidth}
\end{minipage}
\end{table}

This appendix complements the scanner execution setup in Table~\ref{tab:scanner_setup} by summarizing the analysis logic behind each scanner. While the table records the execution mode, input source, and configuration used in our experiments, the descriptions below clarify what evidence each scanner relies on and what type of MCP workflow risk it is designed to surface.

\begin{itemize}
    \item \textbf{Agent-Scan} connects to the MCP server, retrieves tool descriptions, and analyzes them using local rules together with the remote Snyk API~(black-box).

    \item \textbf{A.I.G (static)} analyzes the source code with an LLM Agent, and outputs the risk report.

    \item \textbf{A.I.G (dynamic)} connects to the MCP server, interacts with it by executing commands, automatically interprets the code behavior, generates test payloads, and validates the responses.

    \item \textbf{MCP-Scanner} connects to the MCP server, retrieves tool descriptions, and analyzes them using YARA rules and an LLM.

    \item \textbf{MCPScan} first applies Semgrep-based static source-sink taint analysis to identify potentially risky paths. It then uses an LLM to review tool descriptions, reconstructs data flows, and relies on the LLM to judge the final risk.

    \item \textbf{MCPSafetyScanner} uses multiple agents to perform security checks against the server. The hacker agent executes attack commands, while the auditor agent searches external knowledge bases for attack methods. It then generates a vulnerability report.

    \item \textbf{mcp-gateway} first collects public MCP server metadata from NPM, Smithery, and GitHub. It then estimates the reputation score of the MCP server, connects to the server, and analyzes whether tool description fields contain malicious content.

    \item \textbf{nova-proximity} connects to the MCP server, retrieves tool descriptions, tool prompts, and resources, and analyzes them using NOVA rules and an LLM.

    \item \textbf{mcp-armor} connects to the MCP server, retrieves tool descriptions, prompts, and resources, and analyzes them using regular expressions and a local LLM~(Aira-security/FT-Llama-Prompt-Guard-2).
\end{itemize}

\section{Risk Category Mapping}
\label{appendix:risk-category-mapping}

Section~\ref{sec:result_analysis}
requires a systematic comparison across various scanners to evaluate their performance on MCP servers.
Table~\ref{tab:category_mapping} details the comprehensive mapping rules used to harmonize the heterogeneous output schemas from these different scanners.
For each scanner, we report its native vulnerability labels and their corresponding unified risk categories: \textit{Prompt Injection}, \textit{Command Execution}, \textit{Data Leakage}, and \textit{Other}.

\section{CVE Ground-Truth Dataset}
\label{sec:cve_list}

Section~\ref{sec:validation} uses public CVEs as an external ground truth for evaluating scanner recall on known vulnerable MCP servers.
Table~\ref{tab:cve_ground_truth} lists the CVEs retained after repository matching, vulnerable-version filtering, and risk-scope filtering.
For each CVE, we report the number of affected MCPZoo servers, the normalized risk type used in our paper, the disclosed severity, and the CVSS score.
Because one server may be associated with multiple CVEs, recall is evaluated over applicable scanner--server pairs: for each scanner, an affected server is counted once when the scanner successfully processes it and supports at least one risk category associated with its CVEs.

\begin{table}[h]
\centering
\caption{CVE List Used for CVE-Based Recall Validation in Section~\ref{sec:validation}. The risk type column reports the normalized paper-level risk category.}
\label{tab:cve_ground_truth}
\resizebox{\linewidth}{!}{
\begin{tabular}{l c l c c}
\toprule
\textbf{CVE} & \textbf{Affected Servers} & \textbf{Risk Type} & \textbf{Severity} & \textbf{CVSS} \\
\midrule
CVE-2025-53818~\cite{nvd-cve-2025-53818} & 14 & Command Injection & HIGH & 8.9 \\
CVE-2025-66580~\cite{nvd-cve-2025-66580} & 1 & Code Execution & CRITICAL & 9.6 \\
CVE-2025-68669~\cite{nvd-cve-2025-68669} & 1 & Code Execution & CRITICAL & 9.6 \\
CVE-2026-22793~\cite{nvd-cve-2026-22793} & 1 & Code Execution & CRITICAL & 9.6 \\
CVE-2026-25546~\cite{nvd-cve-2026-25546} & 9 & Command Injection & HIGH & 7.8 \\
CVE-2026-25650~\cite{nvd-cve-2026-25650} & 12 & Credential Leakage & MEDIUM & 6.6 \\
CVE-2026-27825~\cite{nvd-cve-2026-27825} & 8 & Credential Leakage & HIGH & 8.0 \\
CVE-2026-33946~\cite{nvd-cve-2026-33946} & 1 & Code Execution & HIGH & 8.2 \\
CVE-2026-33980~\cite{nvd-cve-2026-33980} & 6 & Prompt Injection & HIGH & 8.3 \\
CVE-2026-39884~\cite{nvd-cve-2026-39884} & 14 & Command Injection & HIGH & 8.3 \\
\bottomrule
\end{tabular}
}
\end{table}

\section{Details for Automated Classification}
\label{appendix:evaluation-prompts}

\subsection{Server Function Classification}
\label{appendix:prompt-server-function}

We classify MCP servers into functional domains based on their primary usage scenarios, as discussed in Section~\ref{sec:server_func}. The classification is performed using a prompt-based LLM annotation.

\begin{promptbox}{Prompt of Server Function Classification}
You are a professional software classification assistant. Given a list of MCP Servers provided by the user (including MD5, serverName, and description), please determine the functional category for each server.

The available categories are as follows:
\begin{itemize}
    \item Finance: Financial and accounting-related functions such as stock trading, market data retrieval, and financial analysis.
    \item E-commerce: E-commerce and online shopping functions.
    \item Medical: Healthcare-related functions, such as medical information queries.
    \item Developer Tools: Development tools, including coding assistance, database operations, and cloud service management.
    \item Web Search: Search tools for information retrieval and knowledge acquisition.
    \item Academical Research: Functions for academic information gathering and research.
    \item Media \& Content Processing: Functions for image, text, and audio editing.
    \item Communication \& Messaging: Email, SMS, and social media message processing.
    \item Utilities: General utility components such as weather information, transit queries, and calculators.
    \item Others: Any other functions not covered by the categories above.
\end{itemize}

\textbf{Output Requirements:}
\begin{enumerate}
    \item You must return a standard JSON list (List of Objects).
    \item The list must include all input items, preserving the original order.
    \item Each object must contain the following fields:
    \begin{itemize}
        \item "MD5": The exact MD5 string provided in the input.
        \item "Server\_Name": The exact serverName provided in the input.
        \item "function\_class": The most appropriate category selected from the list above.
        \item "explanation": A very concise explanation (under 10 words).
    \end{itemize}
    \item Do not output any Markdown formatting (e.g., \texttt{```json}); output only the raw JSON string.
\end{enumerate}
\end{promptbox}

\subsection{Tool Capability Classification}
\label{appendix:prompt-tool-capability}

We categorize MCP tools based on their name, description, and input schema, as discussed in Section~\ref{sec:tool}. The classification is performed using a prompt-based LLM annotation.

\begin{promptbox}{Prompt of Tool Capability Classification}
You are an expert security analyst and software engineer. Your task is to classify a list of MCP (Model Context Protocol) Tools into one or more of the following 7 categories based on their Name, Description, and Input Schema.

\textbf{Classification Categories \& Criteria:}
\begin{itemize}
    \item \textbf{Command/Code Execution:} Tools that execute shell commands, database queries, scripts (Python/JS/Bash), or manage system processes. Keywords: \texttt{shell}, \texttt{bash}, \texttt{cmd}, \texttt{exec}, \texttt{process}, \texttt{terminal}, \texttt{sql}, \texttt{query}, \texttt{eval}.
    \item \textbf{Network Data Sending (Outbound):} Tools that send data from the local environment to external servers (Upload, Email, SMS, Social Media Posting). Keywords: \texttt{post}, \texttt{send}, \texttt{upload}, \texttt{email}, \texttt{push}, \texttt{publish}, \texttt{comment}.
    \item \textbf{Local Data Writing (Modification):} Tools that modify the local filesystem (Write, Append, Delete, Rename, Move files). Keywords: \texttt{write}, \texttt{save}, \texttt{append}, \texttt{delete}, \texttt{remove}, \texttt{mkdir}, \texttt{edit}.
    \item \textbf{Local Info Reading (Privacy Risk):} Tools that read local file contents, take screenshots, or list directory structures. High privacy risk. Keywords: \texttt{read}, \texttt{cat}, \texttt{load}, \texttt{screenshot}, \texttt{list}, \texttt{ls}, \texttt{view}.
    \item \textbf{Network Data Reading (Inbound):} Tools that fetch data from the internet (Browsing, Searching, API Fetching). Risk of prompt injection/poisoning. Keywords: \texttt{search}, \texttt{browse}, \texttt{fetch}, \texttt{get\_url}, \texttt{crawl}, \texttt{scrape}.
    \item \textbf{Prompt Provision:} Tools specifically designed to inject prompts, thinking patterns, or context into the LLM's conversation history. Keywords: \texttt{thinking}, \texttt{reflect}, \texttt{context}, \texttt{memory}, \texttt{prompt}.
    \item \textbf{Others (Utilities):} Pure utility functions like calculators, data format converters, time checks, etc.
\end{itemize}

\textbf{CRITICAL RULES (Batch Processing):}
\begin{enumerate}
    \item \textbf{Category 7 is EXCLUSIVE:} If a tool fits into categories 1-6, you MUST NOT include category 7.
    \item \textbf{Output Format:} You must return a JSON Array.
    \item \textbf{Completeness:} You will receive a list of tools. You must return a result object for EVERY tool in the list, maintaining the \texttt{id}.
    \item \textbf{No Noise:} Do NOT output markdown code blocks. Just the raw JSON string.
\end{enumerate}

\textbf{Output JSON Format:}
\\
{[}
    \{
        "id": "tool\_id\_from\_input",
        "category\_ids": [1, 3],
        "reasoning": "Briefly explain why (max 20 words)."
    \}, ...
{]}

\textbf{User Input Format:}
\\
Please classify the following list of tools according to the rules:

--- TOOL ITEM 1 --- \\
ID: \{tool.id\} \\
Name: \{tool.name\} \\
Description: \{tool.description\} \\
Input Schema: \{tool.schema\_str\}

Output the JSON Array containing classifications for ALL tools above. Remember: Category 7 is exclusive.
\end{promptbox}

\subsection{Validation of Classification}\label{appendix:prompt+validation}

To assess the reliability of the LLM-assisted classification, we perform manual validation on a randomly sampled subset of MCP servers and tools.

Specifically, we randomly sample 100 instances covering all categories. Two authors independently inspect each sample based on descriptions, input/output schemas, and available source code context, and assign ground-truth labels following the same taxonomy. Disagreements are resolved through discussion.

We observe an agreement rate of 96\% between manual annotations and LLM predictions, indicating that the classification is generally reliable. Most discrepancies arise from ambiguous or underspecified descriptions, rather than systematic errors.

Importantly, we do not observe error concentration in specific categories, suggesting limited systematic bias. Overall, while individual misclassifications exist, they do not affect the aggregate-level trends reported in our analysis.

\section{Cross-Scanner Consistency}\label{appendix:consistency}
\begin{figure}[H]
\centering
\includegraphics[width=0.7\linewidth]{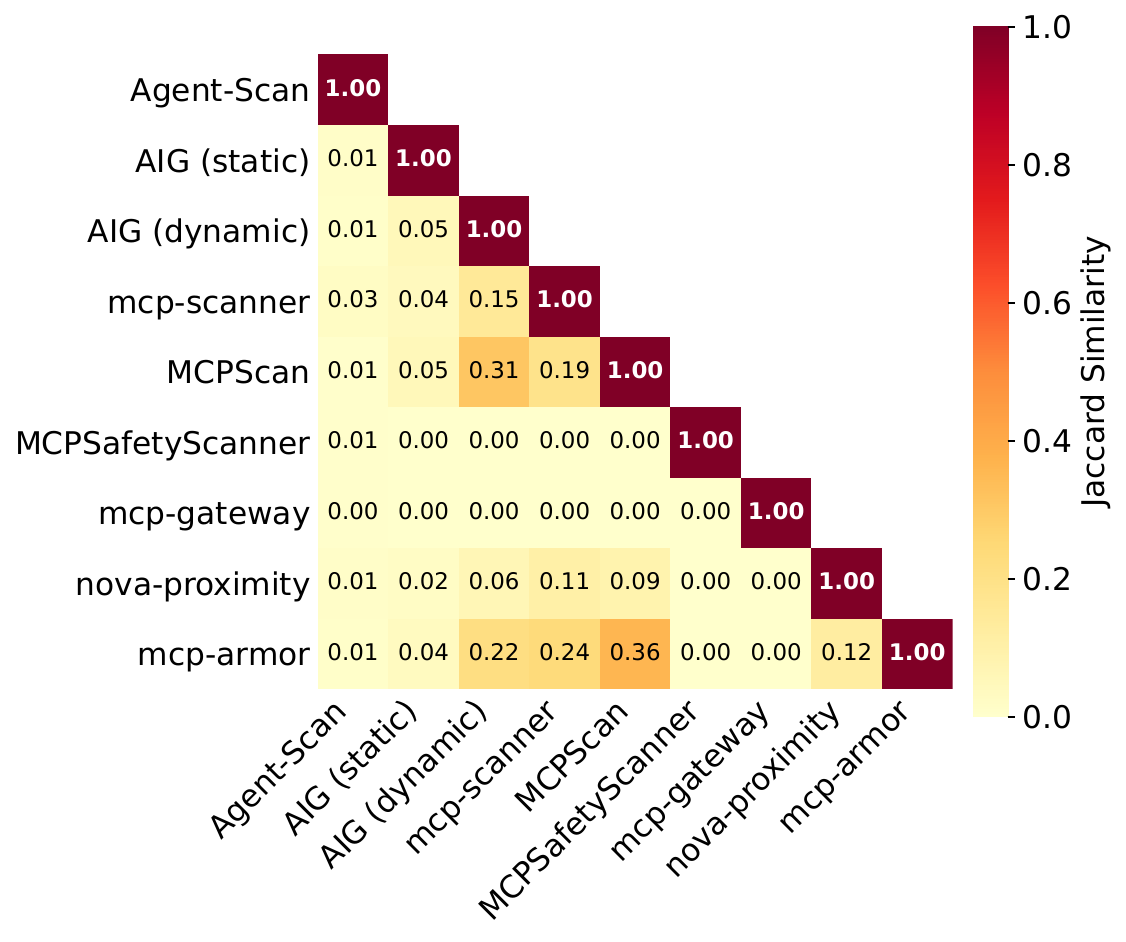}
\caption{Pairwise Jaccard Similarity among Capable Scanners on the \emph{Prompt Injection} Category.}
\Description{Heatmap showing pairwise Jaccard similarity among capable scanners for the Prompt Injection category.}
\label{fig:jaccard_prompt}
\end{figure}

\begin{figure}[H]
\centering
\includegraphics[width=0.7\linewidth]{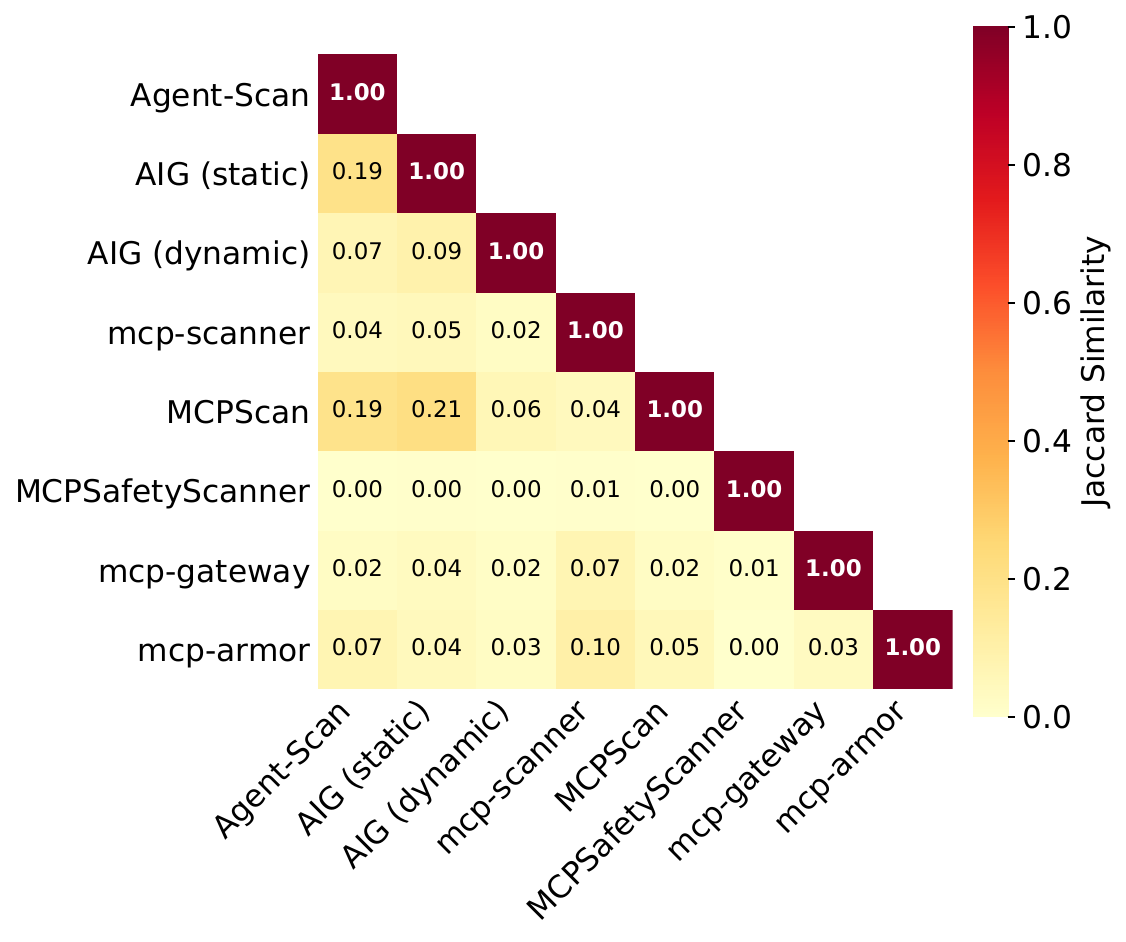}
\caption{Pairwise Jaccard Similarity among Capable Scanners on the \emph{Command Execution} Category.}
\Description{Heatmap showing pairwise Jaccard similarity among capable scanners for the Command Execution category.}
\label{fig:jaccard_code}
\end{figure}

\begin{figure}[H]
\centering
\includegraphics[width=0.7\linewidth]{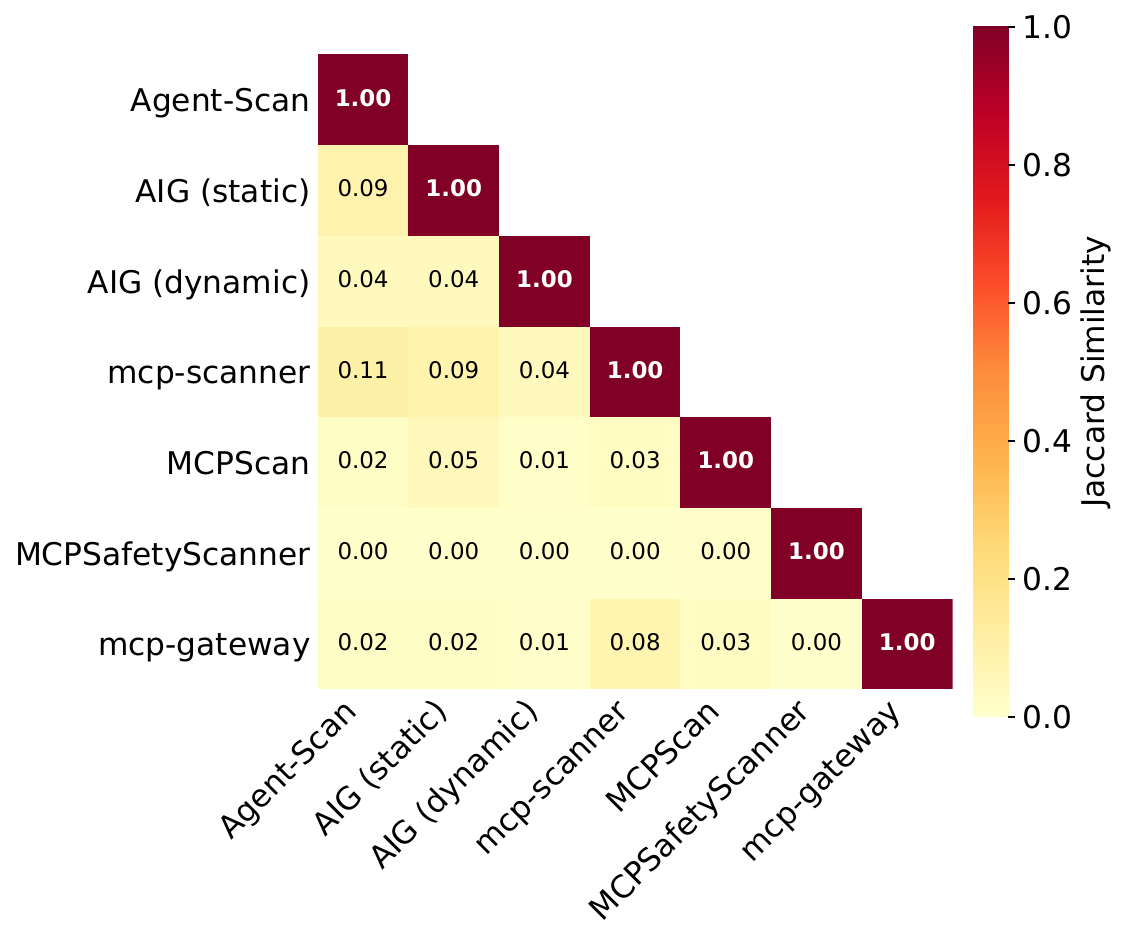}
\caption{Pairwise Jaccard Similarity among Capable Scanners on the \emph{Data Leakage} Category.}
\Description{Heatmap showing pairwise Jaccard similarity among capable scanners for the Data Leakage category.}
\label{fig:jaccard_cred}
\end{figure}

Figures~\ref{fig:jaccard_prompt}, \ref{fig:jaccard_code}, and \ref{fig:jaccard_cred} present the pairwise Jaccard similarity matrices for each vulnerability category, restricted to scanners capable of detecting that category.
Each cell reports the fraction of MCP servers flagged by both tools out of those flagged by either, ranging from 0 (no overlap) to 1 (perfect agreement).
Diagonal entries are always 1.00 by definition.
Across all three categories, inter-scanner agreement remains consistently low, corroborating the global-level finding reported in Section~\ref{sec:validation}.

\end{document}